\documentclass[a4paper,11pt]{article}

\usepackage{geometry}										
\usepackage[english]{babel}									
\usepackage[utf8]{inputenc}									
\usepackage[T1]{fontenc}									

\usepackage{lmodern}										

\usepackage{lineno}

\usepackage{amssymb,amsfonts,amsmath,amsthm}
\usepackage{enumerate}
\usepackage{booktabs}
\usepackage{hyphenat}
\usepackage{bbm}
\usepackage{multirow}
\usepackage{graphicx}
\usepackage{hyperref}
\usepackage{subfigure}
\usepackage{caption}
\usepackage{subcaption}
\usepackage{xcolor}
\usepackage{placeins}
\usepackage{mathtools}
\usepackage{adjustbox}
\usepackage{stfloats} 

\usepackage{booktabs}   
\usepackage{multirow}   
\usepackage{makecell}

\usepackage{soul}
\usepackage{bm}  
\usepackage{tcolorbox}
\makeatletter
\let\oldabs\abs
\def\abs{\@ifstar{\oldabs}{\oldabs*}}
\let\oldnorm\norm
\def\norm{\@ifstar{\oldnorm}{\oldnorm*}}
\makeatother

\newcommand{\dsz}[1]{$\text{DS}_0$}
\newcommand{\dso}[1]{$\text{DS}_1$}
\newcommand{\dsc}[1]{$\text{DS}_2$}
\def\bP{\mathbb P}
\newcommand{\tabref}[1]{Table~\ref{#1}}

\usepackage{booktabs}
\usepackage{multirow}
\usepackage{rotating}
\usepackage{siunitx}

\usepackage{array} 
\newcolumntype{P}[1]{>{\centering\arraybackslash}p{#1}}
\usepackage{authblk}												

\usepackage{rotating}

\usepackage{cases}

\usepackage{xcolor} 

\definecolor{darkred}{rgb}{0.4, 0.0, 0.0} 
\definecolor{darkgreen}{RGB}{177,227,211} 
\definecolor{darkgrey}{RGB}{191,191,191} 
\usepackage{tikz}
\usetikzlibrary{bayesnet}
\usetikzlibrary{calc}
\usetikzlibrary{shapes,fit}

\usetikzlibrary{automata}
\usetikzlibrary {arrows.meta}
\usetikzlibrary{fit,positioning}
\pgfmathsetmacro{\sinOffset}{sin(45)}
\pgfmathsetmacro{\cosOffset}{cos(90)}

\tikzset{  myarrow/.style={    -{Classical TikZ Rightarrow[length=1mm]},}}
\tikzset{  myarrowLong/.style={     draw=darkred,   -{Classical TikZ Rightarrow[length=1mm, width=2mm]},}}
\tikzset{  myarrowDamage/.style={     draw=darkgrey,   -{Classical TikZ Rightarrow[length=1mm, width=2mm]},}}
\tikzset{  myarrowReco/.style={     draw=darkgreen,   -{Classical TikZ Rightarrow[length=1mm, width=2mm]},}}

\newcommand{\flipcurve}[2]{\raisebox{0.75ex}{\scalebox{1.4}[-1]{$#1#2$}}}

\tikzstyle{cont}=[circle,accepting, minimum size = \bulletsize mm, fill = white!100, thick, draw =black!80, node distance = \nodedist mm] 
\tikzstyle{disc}=[circle, minimum size = \bulletsize mm, fill = white!100, thick, draw =black!80, node distance = \nodedist mm]
\tikzstyle{obsv}=[circle, minimum size = \bulletsize mm, fill = red!15, thick, draw =black!80, node distance =\nodedist mm]
\tikzstyle{obsv2}=[circle, minimum size = \bulletsize mm, fill = red!15, thick, draw =black!80, node distance =\nodedist mm]
\tikzstyle{parm}=[circle, minimum size = \bulletsize mm, node distance =\nodedist mm]
\tikzstyle{dist}=[node distance =\nodedist mm]
\tikzstyle{dummy}=[circle, minimum size = \bulletsize mm, node distance =\nodedist mm]
\tikzstyle{connect}=[-latex, thick,>=latex]
\tikzstyle{connect2}=[-latex, thick, dashed, >=latex]
\tikzstyle{shaded}=[draw=black!20,fill=black!15]
\tikzstyle{box2}=[rectangle, draw=black!100, dashed,fill=red!15]
\tikzstyle{box3}=[rectangle, draw=black!100, dashed]
\tikzstyle{box}=[rectangle, draw=black!100, thick, rounded corners]
\tikzstyle{boxblue}=[rectangle, inner sep=2mm,fill=blue!20, thick,rounded corners]
\tikzstyle{nobox}=[rectangle, inner sep=2mm,fill=white!20, thick,rounded corners]

\usepackage{microtype}										

\usepackage{indentfirst}									

\usepackage[numbers]{natbib}
\usepackage{hyperref}
\hypersetup{
pdftitle = {UQ based state-dependent framework for recovery and seismic risk assessment},
pdfauthor = {Chiara Nardin, Stefano Marelli, Marco Broccardo},
pdfsubject = {},
pdfkeywords = {s},
colorlinks = true,                                                                                                                                                                          
allcolors = blue,                                                                                                                                                                          
}

\usepackage[capitalize]{cleveref}								            
\creflabelformat{equation}{#2(#1)#3}								        
\crefrangelabelformat{equation}{#3(#1)#4 to #5(#2)#6}						
\labelcrefformat{subequation}{#2(#1)#3}								        
\labelcrefrangeformat{subequation}{#3(#1)#4 to #5(#2)#6}					

\graphicspath{{./Figures}}

\title{A Generalized Framework for Performance-Based Earthquake Engineering: Integrated Assessment of Structural Reliability and Resilience
}

\author[1,2]{Chiara Nardin\thanks{chiara.nardin@unitn.it, cnardin@ethz.ch}}
\author[2]{Stefano Marelli}
\author[2]{Bruno Sudret}
\author[1]{Marco Broccardo*}
\affil[1]{Department of Civil, Environmental and Mechanical Engineering, University of Trento, Italy}
\affil[2]{Chair of Risk, Safety and Uncertainty Quantification, ETH Z\"{u}rich, Switzerland}

\date{}
\begin{document}
\maketitle
\vspace{-1.5cm}
\begin{abstract}

Assessing structural reliability and resilience under seismic hazard requires accounting not only for damage accumulation but also for the recovery processes that govern post-event structural restoration. In current performance-based earthquake engineering (PBEE) practice, recovery is typically treated as an external, post-processing attribute, while structural performance is modeled through Poissonian exceedance assumptions that implicitly enforce renewability and memorylessness. These assumptions  limit a full integration of structural reliability and resilience assessments, particularly under repeated seismic loading and non-negligible recovery times.

This study proposes a generalized PEER-PBEE formulation in which damage accumulation and recovery are embedded directly into the system dynamics through a continuous-time Markov chain representation. By replacing event-based renewability with state-dependent transitions governed by a single generator matrix, the framework yields a unified description of structural reliability and resilience that remains fully compatible with PBEE performance metrics while relaxing its core Poissonian assumptions. Time-dependent failure probabilities and reliability indices are obtained from the transient dynamics of the generator, enabling robust quantification of cumulative risk. In parallel, a resilience index is introduced to measure the expected fraction of operational time prior to collapse. The metric is expressed in a standardized form analogous to the reliability index, facilitating direct comparison between structural configurations and recovery strategies.

The spectral properties of the generator matrix are exploited to derive both reliability and resilience metrics in a computationally efficient and physically interpretable manner. The approach is first illustrated using a reduced three-state example, which clarifies the underlying mechanisms. It is then applied to two industrial-scale case studies representing common structural archetypes, namely a braced frame and a base-isolated system. The results demonstrate that recovery dynamics can dominate long-term resilience even when conventional reliability metrics show limited sensitivity, highlighting the necessity of generalized PBEE formulations for robust life-cycle performance assessment.

\end{abstract}

\section{Introduction and Motivation}\label{sec:intro}
In performance-based earthquake engineering (PBEE), the paradigm is evolving from an exclusive focus on structural damage to a comprehensive view that includes downtime, functionality loss, and recovery ~\cite{bib:vandelindt2021,bib:terzic2021frec,bib:Cook2022FunctionalRecovery}. This change reflects a fundamental paradigm shift: post-earthquake impacts are determined not only by the extent of physical damage but also by the time required to restore acceptable performance, or \emph{to get back to normal}. For communities, businesses, and critical infrastructure operators, this recovery timeline can be as consequential as the damage itself, often dictating the long-term social, economic, and operational repercussions of an event.

Der Kiureghian~\cite{bib:derkiureghian2005nonergodicity} showed that the Poissonian representation of seismic occurrence, together with the ergodicity assumptions embedded in the PEER-PBEE framework~\cite{bib:cornell2000progress,bib:moehle2004framework,bib:yang2009,bib:leelataviwat1999}, effectively implies immediate recovery and renewal of uncertainties after each event. As a result, the framework is not suited, in its classical form, to represent cumulative damage and finite, non-instantaneous recovery, and therefore does not readily admit the integration of structural resilience within a single coherent framework. Here, the qualifier structural is used deliberately to denote the capacity of a structure to return to its original undamaged configuration while preserving the same level of performance.

Motivated by these observations, the present work generalizes the PEER-PBEE framework by embedding its modular architecture within the broader class of continuous-time Markov chains (CTMCs). In doing so, it extends the framework from seismic risk assessment, here understood as structural reliability under seismic loading, to an integrated treatment of risk and resilience, while preserving its original \textit{modus operandi}. This makes it possible to build upon a highly successful framework while retaining, in transferable form, the knowledge, tools, and analytical methodology developed in the domain of seismic risk assessment. In fact, in this generalized setting, the classical PEER-PBEE formulation is shown to be a special case.

Within this generalized setting, structural resilience acquires a direct representation within PBEE. Whereas the classical formulation expresses performance mainly in terms of mean rates of exceeding prescribed damage or loss thresholds, structural resilience depends on the temporal evolution of repair and restoration after an event~\cite{bib:broccardo2015}. Because recovery may unfold over weeks, months, or years, its explicit representation is essential to capture the persistence of degraded structural performance and the associated broader consequences~\cite{bib:redi2013,bib:terzic2021frec}. This motivates a shift from event-based exceedance measures to a formulation in which performance is described through the stochastic evolution of system states in time, including both damage accumulation and recovery.

Over the past decade, substantial progress has been made in the incorporation of recovery into seismic loss assessment, both through the development of conceptual and probabilistic frameworks and through the implementation of widely used engineering software. In particular, many community- or regional-scale resilience studies rely on recovery functions to estimate repair and recovery times conditional on occupancy class and damage state. A prominent example is the FEMA HAZUS methodology~\cite{bib:fema2020hazus}, which provides a national database of generalized damage, loss, and recovery relationships for large-scale planning and risk assessment, especially within U.S. practice.

Similarly, the Federal Emergency Management Agency (FEMA), with the P-58 project~\cite{bib:fema2018p58_1, bib:fema2018p58_2}, is a U.S. government initiative that developed an extensive taxonomy of fragility and consequence functions based on both post-disaster data and expert judgment. 
Building on FEMA P-58, the Resilience-based Earthquake Design Initiative (REDi)~\cite{bib:redi2013}, developed by U.S. researchers, extended the framework to assess not only repair times but also overall functional recovery of buildings. REDi integrates structural and non structural damage, operational dependencies, and external factors such as permitting and financing delays. It replaces the simplified FEMA P-58 repair schedule with a detailed sequencing model, enabling more realistic predictions of re-occupancy and service restoration timelines. Subsequent refinements, such as the fault tree formulations of Cook et al. (2022)~\cite{bib:Cook2022FunctionalRecovery} and Terzic et al. (2021) \cite{bib:terzic2021frec} have explicitly modeled operational dependencies between structural and non structural systems, using FEMA P-58 database to improve recovery scheduling. 

Building on these developments, the ongoing ATC-138 project~\cite{bib:ATC138_2021}, led by the Applied Technology Council in the United States, provides recovery-based design tools and guidelines for modern buildings. Its methodologies have been implemented in multiple platforms, including the proprietary SP3 platform~\cite{bib:HBRisk2022}, the open-source PELICUN Python package~\cite{bib:Zsarnoczay2024PBE}, and the integrated PACT software~\cite{bib:fema2018p58_2}, which combines FEMA P-58 building inventories with recovery modeling. These software tools translate probabilistic and sequencing-based recovery frameworks into practical applications.

At the same time, analytical studies have formalized recovery modeling, providing the mathematical foundations that underpin these tools. For example, Cassottana et al.~\cite{bib:cassottana2019} introduced a hybrid recovery function, that is a power-law multiplied by an exponential term, providing the first explicit mathematical definition and property set for such functions. Iervolino et al.~\cite{bib:iervolino2015stochastic, bib:iervolino2022holistic} employed stochastic process models, specifically Gamma and Inverse Gaussian processes, to represent the random accumulation of recovery, highlighting their differing probabilistic characteristics and implications for post-earthquake performance modeling.
However, most of the existing PBEE formulations remain limited in their ability to represent the continuous-time evolution of system performance under damage accumulation and recovery mechanisms within a unified probabilistic framework.
In this work, we propose a direct embedding of the recovery process into the PEER-PBEE framework, by replacing the traditional Poisson-based formulation with CTMC model. 

Widely used in the context of classical reliability theory (e.g., \cite{bib:trivedi2017reliability,bib:dhulipala2021ctmc, bib:zeng2021markovReward}), CTMCs provide a natural representation of probabilistic state transitions governed by rates, enabling a time-dependent treatment of both damage accumulation and repair. In the proposed formulation, these rates depend on both the hazard characteristics and the system condition, thereby enabling a continuous-time description of performance evolution. A key feature of the formulation is that it preserves the original modular and hierarchical structure of PBEE, and therefore the same analysis methodology. In particular, the decomposition into hazard and fragility is retained and systematically extended to include recovery, with all components combined within the same analytical architecture. A further key advantage of the CTMC formulation is that the entire stochastic behavior of the system is encoded in a single object, namely the infinitesimal generator matrix. This admits a spectral analysis through which the main system characteristics emerge in a compact and transparent form, including quantities directly related to seismic reliability (i.e., risk) and resilience.
To complement this modeling approach, we introduce a resilience metric $\rho$, along with the classical reliability index $\beta$ \cite{melchers2018structural}. 
The first metric leverages properties of CTMCs to express in closed-form the expected probability of being in a fully operative state given a time horizon. The second instead links directly to the probability of system failure. Thus, by characterizing the entire temporal dynamic evolution, from initial damage to full recovery, our framework supports more accurate resilience evaluations and provides a reliable analytical basis for assessing mitigation and recovery-focused design strategies.

The main contribution of the paper is therefore the development of a CTMC-based generalization of the PEER-PBEE framework that enables the joint, computationally efficient assessment of time-dependent structural reliability and resilience within a unified formulation, while also admitting a spectral analysis through the infinitesimal generator matrix, from which the fundamental characteristics of system behavior emerge in compact form.

The remainder of the paper is organized as follows.
Section~\ref{sec:methodology} introduces the proposed CTMC-based generalization of PEER-PBEE, establishing the theoretical foundations and developing a system-level analysis that exploits CTMC dynamics and spectral properties to capture the temporal evolution of structural performance.
Section~\ref{sec:toy-example} presents an illustrative example that clarifies the mechanics of the framework and highlights the role of spectral properties in shaping system dynamics.
Section~\ref{sec:case-study} applies the methodology to two archetypal industrial-scale seismic mock-ups, demonstrating its practical applicability to complex structural systems.
Finally, Section~\ref{sec:conclusions} discusses the implications and limitations of the proposed approach and outlines directions for future research.

\section{Generalization of PEER-PBEE: \\from a Poissonian to a Markovian framework}\label{sec:methodology}
The classical PEER-PBEE framework operates on the assumption that structural performance under seismic loading can be modeled as a Poisson process, extending a valid representation of earthquake occurrence to the structural response domain. This assumption enables efficient estimation of exceedance probabilities and has been instrumental in the practical implementation of PBEE. However, it also introduces a set of simplifying hypotheses that may constrain the representation of structural performance over extended time horizons.
These assumptions were explicitly identified and critically examined by Der Kiureghian~\cite{bib:derkiureghian2005nonergodicity}, who showed that the Poissonian representation of seismic performance embeds a set of hypotheses that fundamentally limit the role of time and system history in performance assessment. In particular, the framework relies on the following assumptions:
%
\begin{enumerate}[(Hyp~1)] 
    \item \label{hyp:indep_event} Temporal independence of events, implying that the occurrence of one damaging event does not influence the likelihood of future ones;
    \item \label{hyp:stat} Stationarity of the hazard rate $\lambda(im)$, assuming a constant exceedance frequency over time of the seismic intensity measure $im$;
    \item \label{hyp:memory} Memoryless property, implying that the system has no recollection of past events and that the time between successive events is exponentially distributed. 
    \item \label{hyp:renew} Renewability of the system, whereby after each damaging event, the structure is implicitly assumed to return instantaneously to its original, undamaged state. 
\end{enumerate}
Collectively, these assumptions provide a tractable and effective basis for performance assessment, but may reduce the ability to explicitly account for damage accumulation, recovery processes, and path-dependent behavior over long time horizons.
Within this context, the assumption of event renewability becomes particularly restrictive when cumulative damage or irreversible damage, such as collapse, are relevant performance outcomes. 
Structures subject to extreme events (such as collapse or severe functional loss) do not return to their original state after failure. 
Instead, they often require major retrofitting or replacement, which fundamentally alters their physical and performance characteristics. Modeling such transitions as repeatable and memoryless events, as implied by the Poissonian paradigm, can lead to oversimplified and even misleading reliability estimates \cite{bib:derkiureghian2009uncertainties}.

From a PBEE perspective, this limitation manifests as a structural decoupling between damage and recovery: damage accumulation is treated probabilistically within the PBEE workflow, while recovery is typically introduced a posteriori as an external performance modifier, \cite{bib:bruneau2003, bib:cimellaro2010}. As a result, the formulation effectively embeds an implicit and often unrecognized assumption that recovery occurs on a time scale much shorter than that of hazard occurrence, even though empirical evidence indicates that such a separation of time scales is often not observed. Moreover, this separation prevents a unified quantification of reliability and resilience and restricts the ability to compare alternative structural systems or recovery strategies on a consistent probabilistic basis.

To overcome these limitations, we propose a generalized PBEE formulation in which damage accumulation and recovery are embedded directly into the stochastic evolution of the system state. The central idea is to replace the Poissonian framework with a state-based representation of structural performance, in which damage accumulation, irreversible failure, and recovery are treated within a single framework.

To this end, we adopt a CTMC formulation \cite{bib:Ross2014chapter4,bib:trivedi2017reliability} to represent the evolution of the structural state. CTMCs enable explicit modeling of transitions between discrete system states, each of which representing a discrete level of system performance or damage severity, by rate parameters, instead of probabilities. This enables the explicit modeling of cumulative damage, absorbing failure states such as collapse, and recovery processes with state-dependent dynamics, accommodating varying residency times and transition rates. As such, CTMCs provide a natural and rigorous generalization of Poisson-based PBEE, while remaining fully compatible with its probabilistic structure.

\subsection{State-based generalization of PEER-PBEE via CTMC}
To illustrate the modeling objective, consider a classical problem in earthquake engineering: determining the probability that a structure occupies a prescribed undamaged or damage state over a given time interval. In this context, CTMCs provide a more rigorous representation by relaxing the assumption of event renewability (Hyp.~\ref{hyp:renew}). For extreme events, failure does not in general imply immediate restoration of the original system; rather, it leads to a modified system configuration or, in the limiting case, to system replacement.
The proposed generalization of PBEE is achieved by modeling the evolution of structural performance as a stochastic process over a finite set of discrete damage states. Each state represents a distinct level of structural or functional performance, ranging from an undamaged or fully operational condition to a collapsed state. Within this representation, the system evolves continuously in time through transitions between damage states, driven by seismic loading and recovery actions.

Let $ds(t)$ represent the discrete damage state of a structural system at time $t$, and let $\mathcal{D} = \{ds_0, ds_1, \dots, ds_C\}$ denote the finite set of possible system states (e.g., undamaged, partially damaged, collapsed, etc.). 
Because future states are generally unknown, the deterministic variable $ds(t)$ is modeled as a stochastic process $DS(t)$ for all $t > 0$. The probability that the system occupies a particular state $ds_i \in \mathcal{D}$ at time $t$ is denoted as $\pi_i(t) = \bP (DS(t) = ds_i)$, while the full state probability row vector reads $\bm \pi(t) = \left( \pi_0(t), \dots, \pi_C(t)\right)$\footnote{As custom in CTMC literature the probability vector is a row vector, the rest of the vectors will follow the custom vertical notation.}. This formulation replaces event-based renewability with state persistence, allowing past damage to influence future evolution in a physically consistent manner.

The temporal evolution of $\bm{\pi}(t)$ is governed by a continuous-time Markov process, whose dynamics are fully specified by an infinitesimal generator matrix $\bm{Q}$. For $i \neq j$, the element $Q_{ij}$ represents the transition rate from state $ds_i$ to state $ds_j$. Diagonal entries are defined as $Q_{ii} = -\sum_{j \neq i} Q_{ij}$, ensuring that the row sums of $\bm{Q}$ are zero. In the context of PBEE, $\bm Q$ encodes all admissible transitions between damage states and their associated rates, thereby unifying seismic damage accumulation and recovery within a single operator. Unlike classical PBEE exceedance rates, which are constant and state-independent, the transition rates in $\bm Q$ are explicitly conditioned on the current damage state and may reflect non-stationary hazard (i.e, time-variant seismic rates \cite{bib:broccardo2017,bib:iervolino2022holistic}), degradation processes (i.e., time-variant fragility functions \cite{bib:padgett2010,bib:wang2022}), and repair policies (i.e., time-variant recovery rates \cite{bib:dhulipala2021ctmc, bib:lin2017}).

The evolution of the state probability vector $\bm{\pi}(t)$ follows directly from the Chapman–Kolmogorov equations \cite{bib:trivedi2017reliability,bib:Bremaud2020, bib:seabrook2023} for CTMC and leads to the Kolmogorov forward equations:
%
\begin{align}\label{eq:KDF2}
    \dot{\bm{\pi}}(t) &= \bm{\pi}(t) \cdot \bm{Q}, \quad \bm{\pi}(0) = \bm{\pi}_0, 
\end{align}
which define the transient evolution of the system over its service life. 
Conditioning this evolution on non-absorption, or, in other terms, analyzing system behavior prior to collapse, yields time-dependent failure probabilities and reliability indices that naturally generalize those of classical PBEE. In parallel, resilience metrics emerge directly from the same state dynamics by quantifying the expected fraction of time spent in operational states prior to collapse.

This state-based formulation, while relaxing its core Poissonian assumptions, remains consistent with the PEER-PBEE philosophy, since performance is still expressed in probabilistic terms of damage and functionality.  
In particular, it enables (i) cumulative and irreversible damage, (ii) explicit modeling of recovery, and (iii) consistent long-term reliability and resilience assessment within a unified probabilistic framework. The specific structure and interpretation of the generator matrix $\bm Q$, including its decomposition into damage and recovery components, are discussed in the following section.

For completeness, the formal definition of the underlying CTMC and its governing equations is provided in Appendix~\ref{app:ctmc-basics}.

\subsubsection{State-dependent damage transitions: constructing $\mathbf{Q}_{\mathrm{dam}}$ for generalized PBEE}
A key feature of the proposed generalized PBEE formulation is the decomposition of the infinitesimal generator $\bm Q$ into two physically interpretable components: a damage accumulation submatrix and a recovery submatrix. This separation reflects the two competing processes that govern the temporal evolution of structural performance: deterioration due to seismic loading and restoration due to recovery efforts. 

In this paper, we use $\lambda_{ij}$ to denote progressive transitions to more severe states (from $i$ to $j > i$, up to $C$, the collapse state), and $\mu_{ij}$ to denote recovery transitions (from $j$ to $i < j$). Formally, we write:
\begin{align}
\bm Q &= \bm Q_{\mathrm{dam}} + \bm Q_{\mathrm{rec}} =  \nonumber \\ 
&=
\resizebox{0.85\textwidth}{!}{$\begin{bmatrix}
\displaystyle -\sum_{j>0}^{j=\text{C}} \lambda_{0j} & \lambda_{0,1} & \lambda_{0,2} & \dots & \lambda_{0,\text{C}} \\
\mu_{1,0} & \displaystyle -\left(\mu_{1,0} + \sum_{j>1}^{j=\text{C}}\lambda_{1j}\right) & \lambda_{1,2} & \dots & \lambda_{1,\text{C}} \\
\mu_{2,0} & \mu_{2,1} & \displaystyle -\left(\sum_{j=0}^{j=i-1}\mu_{2j}+\sum_{j>2}^{j=\text{C}}\lambda_{2j}\right) & \dots & \lambda_{2,\text{C}} \\
\vdots & \vdots & \ddots & \ddots & \vdots \\
\mu_{\text{C-1},0} & \mu_{\text{C-1},1} & \dots & \mu_{\text{C-1},\text{C-1}} & \displaystyle -\left(\sum_{j=0}^{j=\text{C-1}} \mu_{\text{C-1},j} + \lambda_{\text{C-1},\text{C}}\right) \\
0 & 0 & \dots & 0 & 0 \\
\end{bmatrix}$}
\label{Qtot}
\end{align}
Here, $\bm Q_{\mathrm{dam}} $ is an upper triangular matrix, representing seismic-induced damage accumulation: transitions only occur from less to more damaged states. Conversely, $\bm Q_{\mathrm{rec}} $ is lower triangular matrix, capturing the recovery process: transitions move the system toward improved or less degraded conditions. 
Moreover, the last row of $\bm Q$ corresponds to an absorbing state. Once the system enters this state, no further transitions are possible, thus, representing a terminal, irreversible condition. This is consistent with the physical meaning of collapse in structural systems and underscores the non-renewable character of extreme damage scenarios.

\begin{figure}[!ht]
\centering  
\begin{tikzpicture}

\node[latent,draw=darkgrey,
  fill=darkgrey,             
  fill opacity=0.15,         
  text opacity=1,           
] (d0) [minimum size=.9 cm, label=center:$0$] { }; 
\node[latent,draw=darkgrey,
  fill=darkgrey,             
  fill opacity=0.15,         
  text opacity=1,           
] (d1) [right  =of d0, minimum size=.9 cm, label=center:$1$] { }; 
\node[latent,draw=darkgrey,
  fill=darkgrey,             
  fill opacity=0.15,         
  text opacity=1,           
] (d2) [right  =of d1, minimum size=.9 cm, label=center:$2$] { }; 
\node[]       (d3) [right  =of d2, minimum size=.9 cm, label=center:$\cdots$] { }; 
\node[latent,draw=darkgrey,
  fill=darkgrey,             
  fill opacity=0.15,         
  text opacity=1,           
] (d4) [right  =of d3, minimum size=.9 cm, label=center:$C-1$] { }; 
\node[latent,draw=darkred,
  fill=darkred,             
  fill opacity=0.15,         
  text opacity=1,           
] (d6) [right  =of d4, minimum size=.9 cm, label=center:$C$] { }; 

\node[above=0.6cm of d0, text=black] (lambdaLabel) {$\lambda_{ij}^{\curvearrowright}$};

\node[below=0.6cm of d0, text=black] (muLabel)
{$\mu_{\underset{\mathpalette\flipcurve\curvearrowleft}{ij}}$};
      \draw[myarrowDamage, out=200, in=155, looseness=3] (d0) to (d0);
      \draw[myarrowDamage, out=200, in=155, looseness=3] (d1) to (d1);
      \draw[myarrowDamage, out=200, in=155, looseness=3] (d2) to (d2);
      \draw[myarrowDamage, out=200, in=155, looseness=3] (d4) to (d4);

      \draw[myarrowDamage] (d0) to[bend left=40] node[left] {} (d1);
      \draw[myarrowDamage] (d0) to[bend left=40] node[left] {} (d2);
      \draw[myarrowDamage] (d0) to[bend left=40] node[left] {} (d4);
      \draw[myarrowDamage,draw=darkred] (d0) to[bend left=40] node[left] {} (d6);

      \draw[myarrowDamage] (d1) to[bend left=40] node[left] {} (d2);
      \draw[myarrowDamage] (d1) to[bend left=40] node[left] {} (d4);
      \draw[myarrowDamage,draw=darkred] (d1) to[bend left=40] node[left] {} (d6);

      \draw[myarrowDamage] (d2) to[bend left=40] node[left] {} (d4);
      \draw[myarrowDamage,draw=darkred] (d2) to[bend left=40] node[left] {} (d6);

      \draw[myarrowDamage,draw=darkred] (d4) to[bend left=40] node[left] {} (d6);

      \draw[myarrowDamage] (d1) to[bend left=40] node[left] {} (d0);
      \draw[myarrowDamage] (d2) to[bend left=40] node[left] {} (d0);
      \draw[myarrowDamage] (d2) to[bend left=40] node[left] {} (d1);
      \draw[myarrowDamage] (d4) to[bend left=40] node[left] {} (d0);
      \draw[myarrowDamage] (d4) to[bend left=40] node[left] {} (d1);
      \draw[myarrowDamage] (d4) to[bend left=40] node[left] {} (d2);

    \node[nobox, below=0.25cm of d6, yshift=0.01cm, text=darkred, align=center] (abs) 
    {};

    \node[draw=none, inner sep=0pt, minimum size=0pt, left=0.2cm of d0] (leftpad) {};
    \node[draw=none, inner sep=0pt, minimum size=0pt, right=0.2cm of d4] (rightpad) {};
    \node[draw=none, inner sep=0pt, minimum size=0pt, above=1.50cm of d2] (toppad) {};
    \node[draw=none, inner sep=0pt, minimum size=0pt, below=1.0cm of d2] (bottompad) {};

    \plate[outer sep=0pt,draw=black, loosely dotted] {} {(leftpad)(rightpad)(toppad)(bottompad)(d0)(d1)(d2)(d3)(d4)} { $\text{transient}$ }

\end{tikzpicture}
\caption{State-space diagram. Grey nodes represent transient states whilst $C$ represent the absorbing, ultimate state. Grey arrows: $\lambda_{ij}$ for damage, $\mu_{ij}$ for recovery. Dark red arrows: transitions to the absorbing state. }
\label{fig:diagram-state-full}
\end{figure}
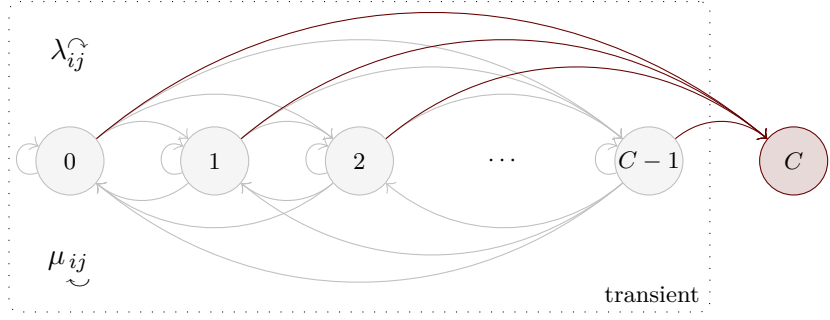

Figure~\ref{fig:diagram-state-full} illustrates the damage state space $\mathcal{D}$ representation of the evolution of the system. In this diagram, the grey-colored nodes—ranging from state $0$ to state $C-1$ (pre-collapse)—denote transient states, meaning states from which the system may exit.
In contrast, the dark-red colored state $C$ represents the absorbing state (collapse), from which the system can neither recover nor receive further damage. 

Accordingly, grey arrows represent transitions between transient states. Specifically, the transitions associated with the rates $\lambda_{ij}$ correspond to damage (moving from state $i$ to a more severe state $j$), while the transitions associated with rates $\mu_{ij}$ correspond to recovery (moving from a damaged state $j$ to a less damaged state $i<j$). The dark-red arrows correspond to transitions into the absorbing state, from which the system cannot exit. Notice that the submatrix corresponding to the transient states forms an irreducible class. This implies that every transient state is reachable from any other.

Notably, modeling seismic damage accumulation via a Markovian framework is not entirely new in earthquake engineering. 
Iervolino et al.~\cite{bib:iervolino2015icasp, bib:iervolino2015eesd} proposed state-dependent Markov models to describe the evolution of damage, capturing cumulative effects and path-dependence in terms of exceedance probabilities. 
Similarly, Andriotis et al.~\cite{bib:andriotis2018} developed a generalized state-dependent framework with richer damage interactions. More recently, Otárola et al.~\cite{bib:otarola2024} introduced a comprehensive discrete-time Markovian framework for multi-hazard life-cycle consequence analysis, which accounts for interactions between instantaneous damage, gradual deterioration (e.g., corrosion), and repair actions through stochastic transition matrices. However, these approaches remain probabilistic in terms of damage exceedance, and do not explicitly represent infinitesimal transition rates. Consequently, while recovery can be modeled as a subsequent discrete step, these frameworks do not allow for the simultaneous, competing dynamics of damage accumulation and restoration.

In contrast, the present work formulates damage accumulation in terms of transition rates $\lambda_{ij}$ between discrete states $ds_i \to ds_j$, which are computed directly from state-dependent PBEE fragilities:
\begin{equation} \label{eq:pbee_lambda_ij}
    \lambda_{ij} = \int_{im} \bP(DS = ds_j | DS = ds_i, im) \left|d\lambda(im) \right| ,
\end{equation}
Here, $ \bP(DS=ds_j \mid DS=ds_i,im) $ represents the state-dependent conditional probability of transitioning from damage state $ ds_i $ to $ ds_j $ given a seismic intensity measure $ im $, while $ \lambda(im) $ is the annual exceedance rate of that intensity measure, i.e., the seismic hazard curve. This construction extends classical PBEE, which assumes stationary exceedance probabilities, by explicitly conditioning on the current damage state and aggregating over all relevant intensity levels, thereby capturing cumulative, path-dependent degradation.

Indeed, in classical PEER-PBEE setup, $\lambda(im)$ is assumed to be stationary over time. That is the exceedance frequency of a given intensity measure is taken as constant across the lifespan of the system. This implies time invariance and temporal independence (Hyp~\ref{hyp:indep_event}-\ref{hyp:stat}): the hazard curve does not change with past events, nor with structural degradation (no cumulative hazard).
A UQ-based computational framework for state-dependent fragility functions, as defined in \eqref{eq:pbee_lambda_ij}, was recently proposed in \cite{bib:nardin2025ress}, highlighting that initial damage can be explicitly accounted for in probabilistic assessments. 

It is worth to notice that the proposed Continuous-Time Markov Chain (CTMC) formulation includes the classical PEER-PBEE framework as a special case. To demonstrate this equivalence, consider a system defined by the state space $\mathcal{D} = \{ds_0, ds_i\}$, where in this case $ds_i$ is a given damage state of interest (e.g., collapse).  In the classical Poissonian approach, the probability of exceeding a performance threshold within a time interval $[0, t]$ is modeled as the probability of ``one or more'' events occurring, given by $P(N(t) \ge 1) = 1 - \exp(-\lambda_{0i}t)$, where $\lambda_{0i}$ is the mean rate of exceedance computed using \eqref{eq:pbee_lambda_ij}.  Within the CTMC framework, this problem is reformulated as a reliability problem where state $i$ is an \textit{absorbing state}. The corresponding transition rate matrix $\mathbf{Q}$ is defined as:

\begin{equation}\label{eq:Q_matrix}
\mathbf{Q} = 
\begin{bmatrix} 
-\lambda_{0i} & \lambda_{0i} \\
0 & 0 
\end{bmatrix}
\end{equation}

Given the initial condition $\boldsymbol{\pi}(0) = (1, 0)$, the solution to the Kolmogorov forward equations yields the state probabilities at time $t$:
\begin{align}
    \pi_0(t) &= \exp(-\lambda_{0i}t) \label{eq:reliability} \\
    \pi_i(t) &= 1 - \exp(-\lambda_{0i}t) \label{eq:failure}
\end{align}
Eq.~\eqref{eq:reliability} represents the system reliability, the probability that the system remains in the unperturbed state throughout the interval $[0, t]$, while Eq.~\eqref{eq:failure} is the system failure probability.  The analytical equivalence between the Markovian absorption probability and the Poissonian solution confirms that the classical PBEE formulation arises directly from the present framework under these restricted conditions. However, framing the problem as a CTMC is more robust for risk analysis; it moves beyond simple counting to provide a natural extension for modeling cumulative damage, serviceability limit states, and recovery processes that are otherwise difficult or not possible to capture within the classical Poissonian assumption.

\subsubsection{State-dependent recovery transitions: constructing $\mathbf{Q}_{\mathrm{rec}}$ for generalized PBEE}
In contrast, the state of the art for recovery is lacking a comparably robust formulation, particularly in industrial contexts. 
Like damage accumulation, recovery is a stochastic process involving transitions between system performance states; however, its dynamics are shaped not only by physical repair mechanisms but also by socioeconomic, logistical, and administrative factors, making it more complex to model consistently. Several frameworks have made strides in addressing this challenge. For example, FEMA P-58~\cite{bib:fema2018p58_1, bib:fema2018p58_2}, its extensions under ATC-138~\cite{bib:ATC138_2021} and HAZUS~\cite{bib:fema2020hazus} provide structured, code-compliant procedures to quantify the repair or functional recovery time of a structure, based on its specific characteristics. Cook \textit{et al.} \cite{bib:Cook2022FunctionalRecovery} provides an overview of the most recent policy shift toward functional recovery. Similarly, tools like the F-RecN + iRe-CoDeS framework~\cite{bib:blagojevic2023frecAndirecodes}, formed by integrating the building-scale F-Rec model \cite{bib:terzic2021frec} with the regional-scale iRe-CoDeS platform—or TREADS, a Python-based simulation environment for modeling building recovery trajectories over time \cite{bib:kourehpaz2022treads, bib:blowes2023}, attempt to represent recovery using fault trees, mobilization delays, and Monte Carlo simulations. While these methods are grounded in practical applications and often supported by empirical data, they are also constrained by their reliance on specific case studies, expert judgment, or regulatory assumptions, which can limit their generalizability across sectors and hazard types.

Despite their practical relevance, most existing recovery models cannot be directly embedded within the PBEE framework, as they are formulated in terms of recovery times or performance trajectories and therefore lack a state-dependent, rate-based representation that can be consistently combined with seismic hazard and damage accumulation. To overcome these limitations, recent efforts advocate for analytical recovery formulations that can both generalize and accommodate empirical insights. For example, Cassottana et al.~\cite{bib:cassottana2019} mathematically defined recovery functions with desirable properties such as boundedness and invariance to time-domain transformations. Their hybrid recovery function family, based on power-law and exponential terms, allows for flexible yet physically meaningful modeling of performance restoration. Stochastic process models like Gamma or Inverse Gaussian processes (e.g., Iervolino et al.~\cite{bib:iervolino2015stochastic, bib:iervolino2022holistic}) further enrich this landscape by enabling time-continuous probabilistic modeling of recovery activities.

Crucially, recovery parameters must often be inferred from post-disaster data or expert elicitation, since controlled experimentation is not feasible. This highlights the importance of a recovery formulation that can fuse empirical data, code-based knowledge, and analytical structures into a unified probabilistic framework.

In analogy with damage modeling, where transition rates $\lambda_{ij}$ are computed via a PEER-PBEE-informed integral over intensity measures (see Eq.~\eqref{eq:pbee_lambda_ij}), the recovery transition rates can be formulated using a complementary structure. For instance, empirical recovery duration distributions, such as those characterized by the REDi framework~\cite{bib:redi2013}, Comerio's safety-tag mapping~\cite{bib:ComerioBlecher2010}, or FEMA-based mobilization models, can be formally mapped to (time-variant) state-to-state recovery rates, analogous to the derivation of hazard rates in classical reliability theory. Alternatively, in the absence of data, analytically defined recovery curves (e.g., from the hybrid family as in~\cite{bib:cassottana2019,bib:iervolino2022holistic}) can be discretized into instantaneous transition probabilities that mirror the damage accumulation side. This dual capability, that is, empirical when data are available, analytical when not, enables the embedding of recovery dynamics directly into the same state-dependent probabilistic architecture that governs seismic damage accumulation \cite{bib:nardin2025ress}, thereby achieving a coherent and scientifically grounded extension of the PEER-PBEE methodology.

\subsection{Structural reliability and resilience assessment under seismic hazard  via CTMC}
Once the infinitesimal generator matrix $\bm Q$ is specified, the CTMC framework enables a unified analysis of the reliability and resilience of the system through both its long-term absorbing behavior and its transient dynamics. On the one hand, absorption analysis quantifies the probability of eventual system collapse. 
On the other hand, transient analysis captures the time-dependent occupancy of non-absorbing states, offering a dynamic picture of functionality and recovery prior to collapse. These two perspectives are complementary: while the absorbing view establishes the ultimate limits of performance, the transient view describes the evolving reliability profile over finite horizons. To further enrich the analysis, spectral methods are introduced. They provide compact descriptions of system dynamics, simplify the computation of mean-time measures, and link damage and recovery rates to the persistence of functionality.

\subsubsection{Absorbing state behavior and long-term probabilities}
\vspace{-2mm}
In general, for a finite-state CTMC, the stationary distribution $\bm{\pi}$ characterizes the long-term probabilistic equilibrium of the system. For an irreducible chain with no absorbing states, $\bm{\pi}$ is unique and represents the steady-state occupancy probabilities, satisfying 
\begin{equation} \label{eq:solution_irredNoabs}
\bm \pi \, \cdot  \bm Q = \bm 0, \quad \sum_i \pi_i = 1,
\end{equation}
as established in classical results from Markov theory \cite{bib:Ross2014chapter4,bib:Bremaud2020}.
However, as in our case, the system can include an absorbing state $C$, representing structural collapse. Because this state is reachable from all operational states and lacks outgoing transitions, the Markov chain is reducible. In this context, the long-term distribution becomes trivial: $\bm \pi = (0, 0, \dots, 0, 1).$ This result simply reflects that, given a sufficiently long time horizon, the system will surely enter the collapsed state. Consequently, the stationary distribution provides no insight into the system's operational life or its recovery dynamics. To quantify performance, we must instead focus on the transient evolution of the state probabilities, which governs the system's reliability and resilience prior to absorption.

\subsubsection{Structural reliability under seismic hazard} \label{sec:reliability}
\vspace{-2 mm}

The behavior of the CTMC is described by the time-dependent row probability vector $\bm \pi(t)$, whose components $\pi_i(t)$ give the probabilities of occupying the corresponding states at time $t$. Assuming that the process starts from state $0$, the initial distribution is $\bm \pi(0)=(1,0,\dots,0)$. The evolution of the state probabilities is governed by the generator matrix $\bm Q$:
\begin{equation}\label{eq:solution_evolution}
\bm \pi(t) = \bm \pi(0) \cdot e^{\bm Q t}.
\end{equation}
When the system includes a single absorbing state, here identified with structural collapse, it is convenient to partition the infinitesimal generator so as to separate the transient dynamics from the absorbing behavior:
\begin{equation} \label{eq:Q_QT}
\bm Q = \begin{bmatrix}
\bm Q_T & \bm a \\
\bm 0^\intercal & 0
\end{bmatrix},
\end{equation}
where $\bm Q_T \in \mathbb{R}^{C \times C}$ is the subgenerator associated with the transient state space $\mathcal{D}_T = \{0,1,\dots,C-1\}$, $\bm a \in \mathbb{R}^{C \times 1}$ collects the transition rates from transient states to the absorbing state, and $\bm 0 \in \mathbb{R}^{C \times 1}$ is the zero vector. Unlike the full generator $\bm Q$, which governs the complete evolution of the CTMC and ultimately transfers all probability mass to the absorbing state, the submatrix $\bm Q_T$ captures the internal damage-evolution dynamics within the transient state space. In the infinite-horizon limit, the transient-state probabilities vanish and collapse occurs, as expected for an absorbing CTMC. Graphically, the term $\bm Q_T$ represents the grey-colored arrows in Figure~\ref{fig:diagram-state-full}, while $\bm a$ represents the red ones. The bottom row of zeros reflects the fact that, once entered, the absorbing state is terminal and cannot be exited. Throughout, system failure is identified with absorption into the collapse state of the CTMC.

This decomposition induces the corresponding block structure in the full transition probability matrix $\bm P(t)=e^{\bm Q t}$ (see the Appendix~\ref{app:ctmc-ptt} for formal definition). In particular, the transient-to-transient block satisfies $\bm P_{TT}(t)=e^{\bm Q_T t}.$ Accordingly, for $i,j \in \mathcal{D}_T$, the entry $[\bm P_{TT}(t)]_{ij}$ gives the probability that the process is in transient state $ds_j$ at time $t$, given that it started in transient state $ds_i$. Since part of the probability mass may already have been transferred to the absorbing state, $\bm P_{TT}(t)$ is, in general, sub-stochastic. 

We now introduce $\mathcal{T}$, the random variable denoting the time to the absorbing state $ds_C$ (i.e., time to collapse). Let $\bm \pi_T(t)$ denote the row probability vector over the transient states at time $t$. Its evolution is governed by the transient generator $\bm Q_T$, namely $\bm \pi_T(t)=\bm \pi_T(0)\,e^{\bm Q_T t},$
with $\bm \pi_T(0)= \bm e_0^\intercal=(1,0,\dots,0)$ and \(\bm e_0\) denoting the canonical basis column vector associated with state \(ds_0\). The cumulative distribution function of $\mathcal{T}$ is then
\begin{equation} \label{eq:failure_F_tau}
    F_\mathcal{T}(t) = \bP(\mathcal{T} \le t) = 1 - S_{\mathcal{T}}(t)
    = 1 - \bm \pi_T(0) \cdot e^{\bm Q_T t} \cdot \bm 1_T
    = 1 - \bm \pi_T(t)\cdot\bm 1_T,
\end{equation}
where $\bm 1_T$ is the column vector of ones with dimension equal to the number of transient states, and
$
S_{\mathcal{T}}(t)=\bm \pi_T(0) \cdot e^{\bm Q_T t} \cdot \bm 1_T=\bm \pi_T(t)\cdot\bm 1_T
$ is the survival function.

Consequently, $F_\mathcal{T}(t)$ is the probability that collapse has occurred during the time interval $[0,t]$ \cite{bib:trivedi2017reliability,bib:Ross2014chapter4,bib:Bremaud2020}. Notice that for the given initial condition this is completly governed by the first raw of $\bm P_{TT}(t)$. Within a PBEE context, this quantity represents the probability of collapse over a prescribed time horizon (e.g., the probability of collapse within 50 years) and reduces to Eq.~\eqref{eq:failure} in the special case of a two-state system.

This probabilistic characterization admits a direct interpretation in terms of structural reliability. Using the generalized reliability index \cite{bib:ditlevsen1979}, one may write
\begin{equation}\label{eq:beta}
\beta_t=  -\Phi^{-1}(F_\mathcal{T}(t)),
\end{equation}
where $\Phi(\cdot)$ denotes the standard normal cumulative distribution function. Eq.~\eqref{eq:beta} provides the structural reliability measure associated with the time interval $[0,t]$.

\subsubsection{Structural resilience under seismic hazard} \label{sec:resilience}
\vspace{-2 mm}
The reliability index $\beta_t$ provides a scalar representation of the probability that the system has not collapsed by time $t$. However, it does not distinguish between a system that, prior to collapse, spends most of its time in the fully operational state and one that stays in intermediate damage states. Structural resilience requires finer information than survival alone, namely the extent to which structural performance is preserved over time. This motivates a shift from the time-to-collapse $\mathcal{T}$ to the occupation time of the undamaged state $ds_0$ prior to collapse. First consider the occupation time of the undamaged state \(ds_0\) over the interval \([0,t]\), defined as
\begin{equation}\label{eq:occupation_time_0}
A_0(t):=\int_0^t \mathbf{1}_{\{DS(s)=ds_0\}}\,ds,
\end{equation}
where  \(\mathbf{1}_{\{\cdot\}}\) is the indicator function. The random variable \(A_0(t)\) represents the total time spent in the undamaged state during \([0,t]\).
Assuming that the process starts from state \(ds_0\), i.e. $\bm \pi_T(0)=\bm e_0^\intercal$, a quantity of particular interest is the expected occupation time of \(ds_0\) over \([0,t]\), conditional on no collapse having occurred by time \(t\). This is given by
\begin{equation}\label{eq:conditional_mean_occupation}
\mathbb{E}[A_0(t)\mid \mathcal T>t]
=
\frac{\mathbb{E}[A_0(t)\mathbf{1}_{\{\mathcal T>t\}}]}{\mathbb{P}(\mathcal T>t)}.
\end{equation}
Let define
$\bm D_0:=\bm e_0\bm e_0^\intercal$,
so that \(\bm D_0\) selects the undamaged state. It can be shown (See Appendix~\ref{app:ctmc-a0}) that the numerator can be written as
\begin{equation}\label{eq:conditional_mean_occupation_numerator}
\mathbb{E}[A_0(t)\mathbf{1}_{\{\mathcal T>t\}}]
=
\int_0^t
\bm e_0^\intercal\,e^{\bm Q_T s}\,\bm D_0\,e^{\bm Q_T(t-s)}\,\bm 1_T\,ds.
\end{equation}
Introducing the matrix-valued kernel
\begin{equation}\label{eq:M0_def}
\bm M_0(t):=
\int_0^t
e^{\bm Q_T s}\,\bm D_0\,e^{\bm Q_T(t-s)}\,ds,
\end{equation}
one obtains the compact representation
\begin{equation}\label{eq:conditional_mean_occupation_compact}
\mathbb{E}[A_0(t)\mid \mathcal T>t]
=
\frac{ \bm e_0^\intercal\,\bm M_0(t)\,\bm 1_T}
{S_{\mathcal T}(t)}.
\end{equation}
Normalising by $t$ yields the conditional mean occupation fraction of the undamaged state,
\begin{equation}\label{eq:Q0_def}
\mathcal{Q}_0(t)
= \frac{\mathbb{E}[A_0(t)\mid \mathcal T>t]}{t}
= \frac{\bm e_0^\intercal\,\bm M_0(t)\,\bm 1_T}{t\,S_{\mathcal T}(t)},
\end{equation}
which represents the expected fraction of $[0,t]$ during which the system resides in the undamaged state, given that it has not collapsed by time $t$. Equivalently, $\mathcal{Q}_0(t)$ may be interpreted as the probability that a time instant drawn uniformly at random from $[0,t]$ falls in a period during which the system occupies state $ds_0$, conditional on survival up to $t$. A value of $\mathcal{Q}_0(t)$ close to unity indicates a system that spends most of its lifetime in the undamaged state, whereas a value close to zero indicates a system persistently in a damaged condition.

Next, let $\mathcal{R}_0$ be the complement of $\mathcal{Q}_0$
\begin{equation} \label{eq:conditional_mean_occupation_fraction}
\mathcal{R}_0(t) = 1 - \mathcal{Q}_0(t).    
\end{equation}
Then, $\mathcal{R}_0$ represents the mean fraction of $[0,t]$ spent in the damaged states, conditional on survival. Since $\mathcal{R}_0 \in [0,1]$, in analogy with the reliability index $\beta$ in Eq.~\eqref{eq:beta}, we introduce the resilience index
\begin{equation}\label{eq:rho}
\rho_t = - \Phi^{-1}(\mathcal{R}_0(t)),
\end{equation}
which maps $\mathcal{R}_0$ one-to-one into the standard normal space.

\subsubsection{Spectral analysis}
As established in Sections~\ref{sec:reliability} and~\ref{sec:resilience}, both seismic reliability and resilience derive from the transient dynamics encoded in $e^{\bm Q_T t}$: reliability is derived from the failure probability $F_{\mathcal{T}}(t)$ in Eq.~\eqref{eq:failure_F_tau}, while resilience is derived from $\mathcal{R}_0(t)$ in Eq.~\eqref{eq:conditional_mean_occupation_fraction}. The spectral decomposition of $\bm Q_T$ offers a complementary perspective: it decomposes $e^{\bm Q_T t}$ into a superposition of exponential modes, each associated with a distinct timescale, and thereby reveals how damage accumulation and recovery rates jointly govern the long-term evolution of system performance.

\paragraph{Eigenmode expansion of $e^{\bm Q_T t}$.}
Assume that the transient generator $\bm Q_T$ is diagonalizable. Since $\bm Q_T$ is a stable matrix (all eigenvalues have strictly negative real parts), the matrix exponential $e^{\bm Q_T t}$ admits the eigenmode expansion
\begin{equation}\label{eq:eigenmode_decomposition}
    e^{\bm Q_T t} = \sum_{k=1}^{C} e^{\sigma_k t}\, \bm w_k \bm \nu_k,
\end{equation}
where $\sigma_k$ are the eigenvalues of $\bm Q_T$, $\bm w_k$ are the corresponding right eigenvectors (columns), and $\bm \nu_k$ are the corresponding left eigenvectors (rows).
The state distribution at time $t$, starting from $\bm \pi_T(0)$, is therefore
\begin{equation}
    \bm \pi_T(t) = \bm \pi_T(0)\, e^{\bm Q_T t} 
    = \sum_{k=1}^{C} c_k\, e^{\sigma_k t}\, \bm \nu_k,
\end{equation}
where $c_k = \bm \pi_T(0)\,\bm w_k$ are the modal coefficients determined by projecting the initial distribution onto the eigenbasis. Each mode decays at its own rate $|\sigma_k|$: modes with large $|\text{Re}(\sigma_k)|$ decay rapidly and govern short-term dynamics, while the mode closest to zero dominates at longer timescales.

We sort eigenvalues by decreasing real part: $0 > \text{Re}(\sigma_1) \geq \text{Re}(\sigma_2) \geq \cdots \geq \text{Re}(\sigma_C)$. The dominant eigenvalue $\sigma_1$ — the one with real part closest to zero — controls the slowest-decaying mode and sets the characteristic timescale $1/|\sigma_1|$ over which probability mass flows toward absorption. A smaller $|\sigma_1|$ corresponds to slower degradation and longer persistence in partially damaged states; a larger value indicates faster progression toward collapse. The spectral gap $|\sigma_2 - \sigma_1|$ controls how quickly higher-order modes become negligible relative to the dominant one. When the spectral gap is large, the conditional distribution over $\mathcal{D}_T$ converges rapidly to the quasi-stationary distribution (QSD), defined next.

\paragraph{Quasi-stationary distribution.}
Assume that $\bm Q_T$ is irreducible. Let $\bm \nu$ and $\bm w$ be the left and right eigenvectors of $\bm Q_T$ associated with the dominant eigenvalue $\sigma_1 < 0$, normalized so that
\begin{equation}\label{eq:qsd_eigenvectors_main}
\bm \nu\, \bm Q_T = \sigma_1\, \bm \nu,
\qquad
\bm Q_T\, \bm w = \sigma_1\, \bm w,
\qquad
\bm \nu\, \bm 1_T = 1,
\qquad
\bm \nu\, \bm w = 1.
\end{equation}
The normalized transient distribution
\begin{equation}\label{eq:tilde_pi_qsd_main}
\widetilde{\bm \pi}_T(t) := 
\frac{\bm \pi_T(t)}{\bm \pi_T(t)\,\bm 1_T}
\end{equation}
converges, as $t\to\infty$, to the QSD $\bm q = \bm \nu$, independently of the initial condition. In particular, the conditional probability of occupying the undamaged state satisfies
\begin{equation}\label{eq:tilde_pi0_limit_main}
\widetilde{\pi}_0(t) \to q_0 = \nu_0, \qquad t \to \infty.
\end{equation}
The QSD thus describes the expected distribution of damage states, conditioned on survival, once transient modes have decayed. $q_0$ measures how much of the conditional mass concentrates on the fully operational state.

\paragraph{Asymptotic approximation of $\mathcal{R}_0$.}
We now derive the long-time behavior of $\mathcal{R}_0(t)$ using the spectral properties above. Standard spectral asymptotics applied to the survival function yield
\begin{equation}\label{eq:survival_asymptotic_main}
S_{\mathcal{T}}(t) = \bm \pi_T(0)\,e^{\bm Q_T t}\,\bm 1_T 
\approx w_0\, e^{\sigma_1 t}, \qquad t \to \infty,
\end{equation}
where $w_0 = \bm e_0^\intercal\,\bm w$ is the initial-state component of the dominant right eigenvector. 
To obtain the asymptotics of $\mathcal{R}_0$, we expand the kernel $\bm M_0(t)$ defined in Eq.~\eqref{eq:M0_def}. Substituting the dominant-mode approximation $e^{\bm Q_T s} \approx e^{\sigma_1 s}\,\bm w\,\bm\nu$ into the integrand gives
\begin{equation}\notag
\bm e_0^\intercal e^{\bm Q_T s} \bm D_0 e^{\bm Q_T(t-s)}
\bm 1_T \approx e^{\sigma_1 s} (\bm e_0^\intercal \bm w)
\cdot (\bm\nu\,\bm D_0\,\bm w) \cdot e^{\sigma_1(t-s)} (\bm\nu\,\bm 1_T) = w_0^2\,\nu_0\,e^{\sigma_1 t},
\end{equation}
where we used $ \bm\nu\,\bm D_0 = \nu_0\,\bm e_0^\intercal$, $\bm e_0^\intercal \bm w = w_0$, $\bm\nu\,\bm 1_T = 1$. The scalar $\nu_0 = q_0 $ is precisely the QSD weight of the undamaged state. Integrating over $[0,t]$ and dividing by $t\,S_{\mathcal{T}}(t) \sim t\,w_0\,e^{\sigma_1 t}$, the exponentials cancel and one obtains
\begin{equation}\label{eq:R0_asymptotic_main}
\mathcal{R}_0(t) = 1 - \frac{\bm e_0^\intercal\,\bm M_0(t)\,\bm 1_T}{t\,S_{\mathcal{T}}(t)} \approx 1 - \frac{w_0^2\,q_0\,t\,e^{\sigma_1 t}}{t\,w_0\,e^{\sigma_1 t}} = 1 - w_0\,q_0, \qquad t\to\infty.
\end{equation}
Thus
\begin{equation}\label{eq:R0_limit_complement_main}
\mathcal{R}_0^{\infty} = \lim_{t\to\infty}\mathcal{R}_0(t) =1 - w_0\,q_0.
\end{equation}
\color{black}
This compact expression links the macroscopic resilience metric to two spectral quantities: $q_0 = \nu_0$, the conditional probability of occupying the undamaged state given non-absorption (capturing \emph{where} the system tends to reside), and $w_0 = \bm e_0^\intercal \bm w$, the weight of the initial state on the dominant mode (capturing the initial \emph{excitation} of the slow dynamics). The two roles are 
complementary: $q_0$ reflects the long-run configuration of surviving trajectories, while $w_0$ controls the amplitude of the dominant decay toward absorption. Neither alone suffices to characterize resilience: a system may have large $q_0$ but collapse quickly (small $w_0$), or persist long but mostly in damaged states (small $q_0$).

The approximation in Eq.~\eqref{eq:R0_limit_complement_main} is accurate whenever the spectral gap $|\sigma_2 - \sigma_1|$ is sufficiently large relative to the timescale of interest, so that higher-order modes have decayed and the dominant-mode description is valid. When no clear spectral gap exists, the full expansion in Eq.~\eqref{eq:eigenmode_decomposition} must be retained. A detailed verification of the spectral gap and the accuracy of the approximation is carried out in the industrial-scale application of Section~\ref{sec:case-study}.

\section{Application to a simple three-state analytical model}\label{sec:toy-example}
This section presents a simple explanatory example to illustrate the notation and key concepts introduced in Section~\ref{sec:methodology}. The example also provides physical insight into how reliability and resilience metrics arise from the stochastic description of damage and recovery.

We consider a system with three damage states, $\mathcal{D} = \{0, 1, 2\}$ representing undamaged, damaged but recoverable, and collapsed structural conditions. The collapsed state $ds_2$ is modeled as absorbing, \textit{i.e.} recovery is no longer possible. This assumption is consistent with many structural engineering applications, such as heritage buildings, the collapse of which typically implies irreversible loss.

\begin{figure}[!ht]
    \centering
    \includegraphics[width=0.50\linewidth]{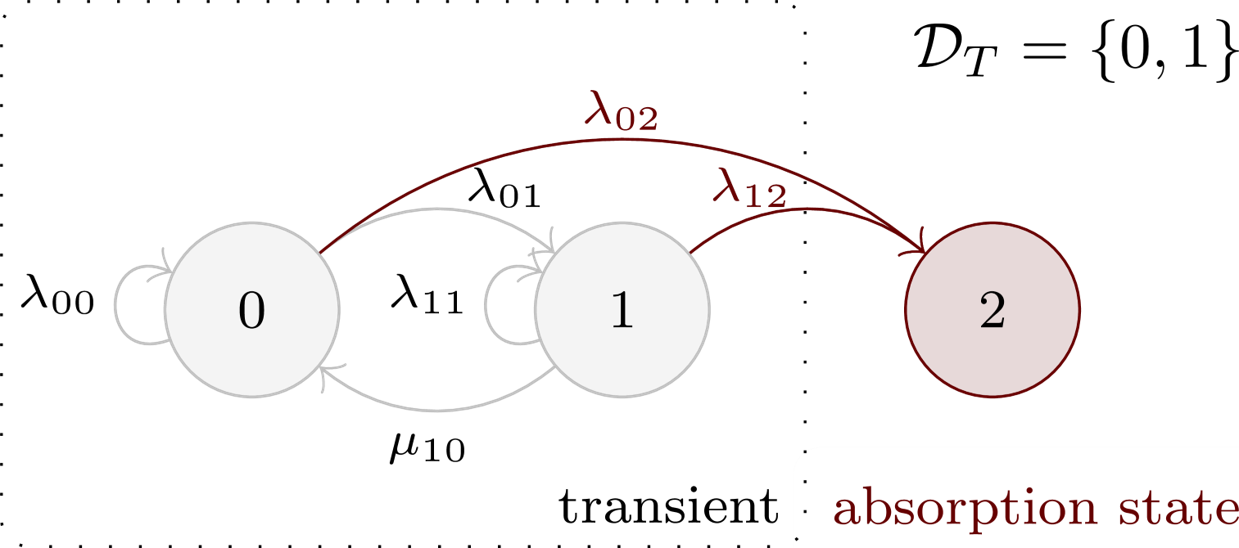}
    \caption{Three-state CTMC representing damage, recovery, and collapse. States $0$, $1$, and $2$ denote the undamaged, damaged but recoverable, and collapsed conditions, respectively. Seismic damage transitions are governed by rates $\lambda_{ij}$, while recovery is described by rate $\mu_{10}$. State $2$ is absorbing.}
    \label{fig:toy-example}
\end{figure}

Figure~\ref{fig:toy-example} illustrates the corresponding CTMC diagram. Transitions driven by seismic damage are governed by rates $\lambda_{ij}$, while recovery processes are described by rates $\mu_{ij}$. Because collapse is irreversible, recovery is only permitted from the damaged state to the undamaged state through the rate $\mu_{10}$. Consequently, the transient state space reduces to $\mathcal{D}_T = \{0, 1\}$, which represents the set of operational and recoverable conditions.
To highlight the role of the different physical mechanisms governing system evolution, we introduce a dimensionless parametrization using the recovery rate $\mu_{10}$ as reference. We define the dimensionless time $\tilde{t} = \mu_{10}\,t$ together with the dimensionless parameters
\begin{equation}\label{eq:parameters}
    \alpha = \frac{\lambda_{01}}{\mu_{10}},
    \qquad
    \gamma = \frac{\lambda_{02}}{\mu_{10}},
    \qquad
    \varepsilon = \frac{\lambda_{12}}{\lambda_{01}}.
\end{equation}

Here, $\alpha$ represents the rate of damage accumulation toward recoverable states relative to the recovery rate; $\gamma$ quantifies the tendency to transition directly to collapse relative to recovery; and $\varepsilon$ controls the conditional probability of progressing to collapse once damage has initiated, relative to the initial damage rate. Together, these parameters describe the competition between three fundamental mechanisms: $\alpha$ governs the dynamics within the recoverable regime; $\gamma$ controls irreversible loss of functionality; and $\varepsilon$ modulates damage progression once degradation has begun.

With this parametrization, the dimensional generator matrix $\bm Q$ factors as
\begin{equation}\label{eq:Q_param}
    \bm Q = \mu_{10}\,\widetilde{\bm Q}(\alpha,\gamma,\varepsilon),
    \qquad
    \widetilde{\bm Q}(\alpha,\gamma,\varepsilon) =
    \begin{pmatrix}
        -(\alpha + \gamma) & \alpha          & \gamma       \\
        1                  & -(1+\alpha\varepsilon) & \alpha\varepsilon \\
        0                  & 0               & 0
    \end{pmatrix},
\end{equation}
where $\bm Q$ carries units of $[\text{year}]^{-1}$ through the prefactor $\mu_{10}$, and $\widetilde{\bm Q}$ is dimensionless. Equivalently, in the rescaled time $\tilde{t} = \mu_{10} t$, the Kolmogorov equations become $\dot{\bm\pi} = \bm\pi\,\widetilde{\bm Q},$ so that the entire time evolution depends only on the three dimensionless parameters $(\alpha, \gamma, \varepsilon)$.

The adopted parameter ranges are grounded in engineering practice and reflect plausible combinations of seismic demand and recovery capacity. As a reference, a direct-collapse rate $\lambda_{02} \simeq 10^{-3}\,[\text{year}]^{-1}$ corresponds to a collapse exceedance probability of approximately 10\% over a 50-year exposure period, consistent with commonly used seismic performance targets. When combined with characteristic recovery times ranging from hours (e.g., systems with rapid re-centering or base-isolation mechanisms) to several years or decades (e.g., conventional structures requiring extensive repair), the resulting dimensionless ratios span the ranges reported in Table~\ref{tab:parameters}.

\begin{table}[!ht]
  \centering
  \caption{Adimensionalized parameters governing damage progression, recovery, and collapse in the three-state CTMC. 
  }
  \label{tab:parameters}
    \scalebox{0.85}{
  \begin{tabular}{llp{7cm}c}
    \toprule
    \textbf{Symbol} & \textbf{Definition} & \textbf{Physical Interpretation} & \textbf{Range} \\
    \midrule
    $\alpha = \lambda_{01} / \mu_{10}$ 
        & Damage-to-recovery ratio 
        & Quantifies how frequently the system transitions from the undamaged state to a damaged but recoverable state, relative to the characteristic recovery rate. 
        & $[10^{-6},\,10^{-1}]$ \\[4pt]

    $\gamma =  \lambda_{02} / \mu_{10}$ 
        & Collapse-to-recovery ratio
        & Measures the tendency to transition directly to collapse compared to the recovery capability of the system. 
        & $[10^{-6},\,10^{-1}]$ \\[4pt]
& &  & \\
    $\varepsilon = \lambda_{12} / \lambda_{01}$ 
        & Damage escalation ratio
        & Describes the propensity of a damaged system to progress toward collapse relative to the onset of damage from the undamaged state. 
        & $[10^{-3},\,10^{2}]$ \\
    \bottomrule
  \end{tabular}}
\vspace{0.35em} 
\parbox{1.5\linewidth}{\scriptsize 
\textit{Note:} The recovery rate $\mu_{10}$ is fixed to $1\,\mathrm{year}^{-1}$ and provides the reference time scale for dimensionless normalization.
}
\end{table}

\begin{figure}[!ht]
    \centering
    \includegraphics[width=0.70\linewidth]{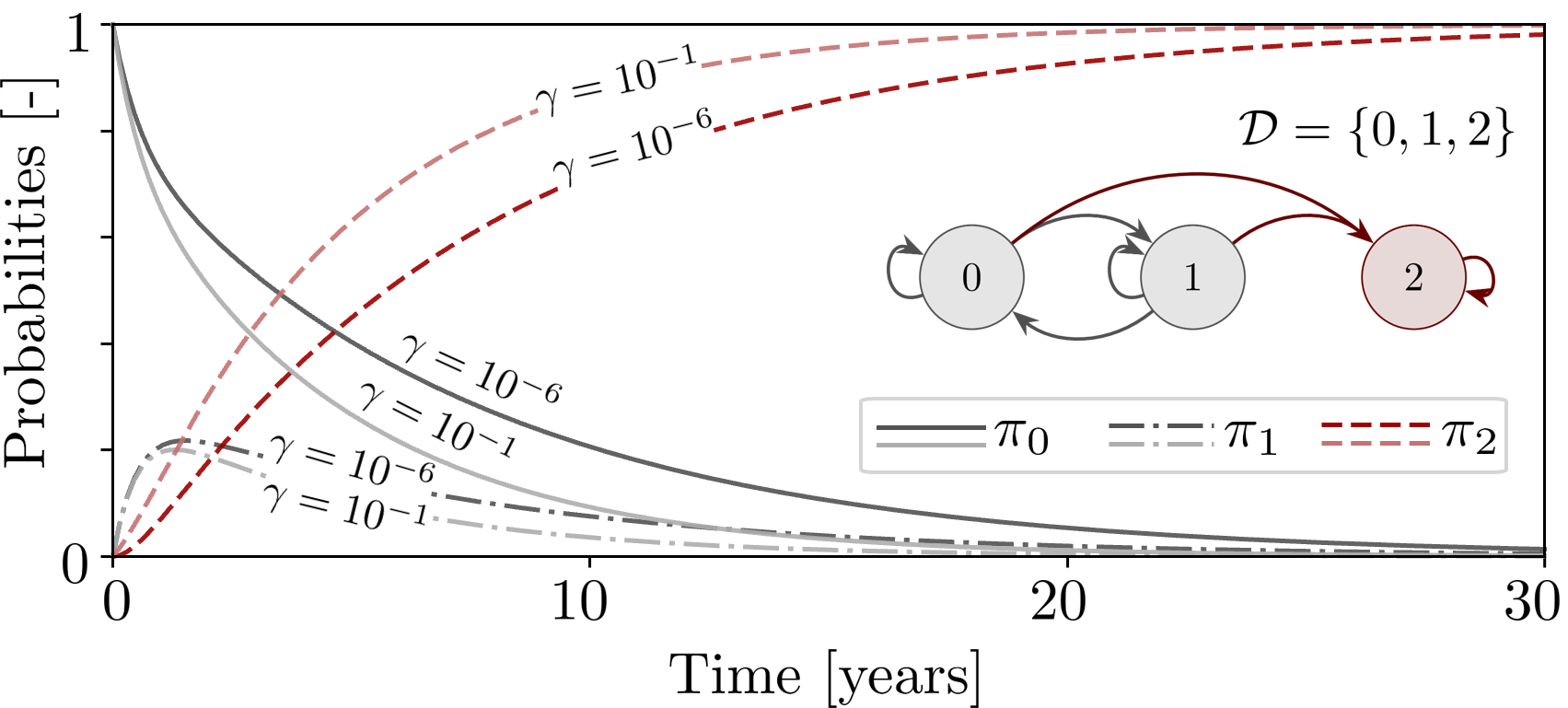}
    \caption{Time evolution of state probabilities for the three-state CTMC model for the two extreme $\gamma$'s value of Table~\ref{tab:parameters}. The system is initialized in the undamaged state, $\bm{\pi}(0) = (1,0,0)$, with fixed $\alpha =0.5$, $\varepsilon=2.0$, and $\mu_{10}=1$.}
    \label{fig:toy-example-long-term}
\end{figure}

\begin{figure}
    \centering
    \includegraphics[width=1.0\linewidth]{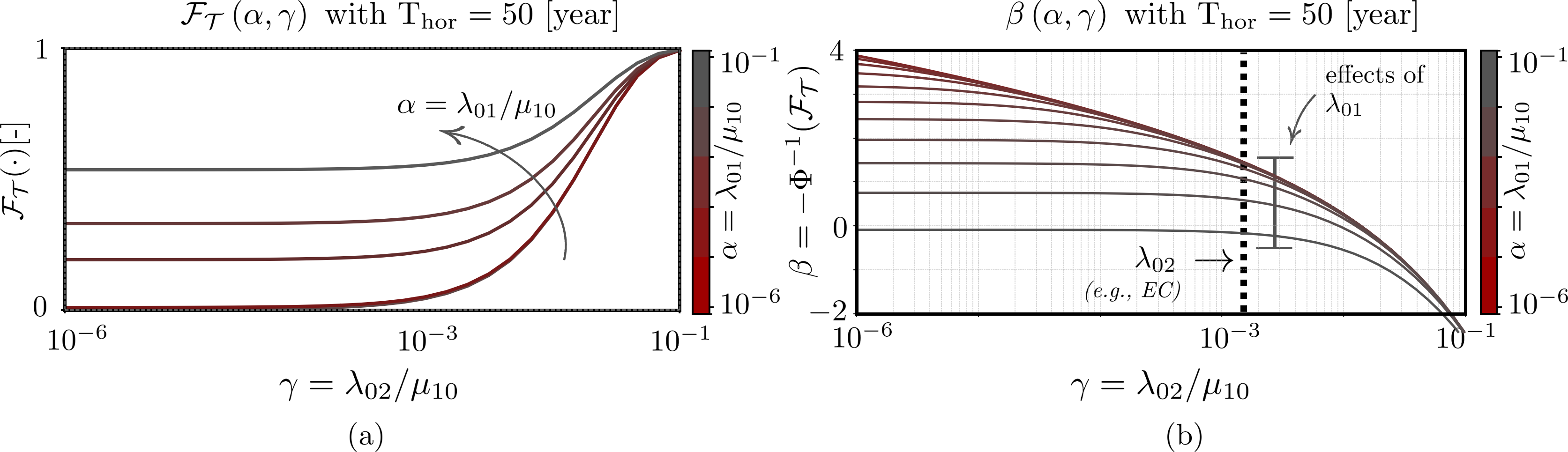}
    \caption{(a) Failure probability $F_{\mathcal{T}}(\cdot)$ evaluated at the time horizon $T_{\text{hor}}$. (b) Corresponding reliability index $\beta$, showing a monotonic decrease when either collapse transitions become more frequent or recovery becomes less effective. The dashed vertical line marks a reference collapse rate $\lambda_{02} \approx 1/475\,\text{[year]}^{-1}$, representative of seismic design prescriptions. Curves illustrate how increasing $\alpha$ reduces reliability for a fixed collapse hazard.} 
    \label{fig:toy-beta2}
\end{figure}

Fixing $\alpha,\varepsilon$, and $\mu_{10}$, the temporal evolution of the state probabilities follows directly from the Kolmogorov forward equations in~\eqref{eq:solution_evolution}, see Appendix~\ref{app:ctmc-basics} for further details. Figure~\ref{fig:toy-example-long-term} illustrates this evolution for increasing values of $\gamma$, assuming the initial pristine condition $\bm \pi(0) = (1.0, 0.0, 0.0)$. As $\gamma$ increases, the probability mass progressively shifts toward the absorbing state. The probabilities associated with the transient states $ds_0$ and $ds_1$ decay accordingly, reflecting the growing likelihood of irreversible collapse. Physically, larger values of $\gamma$ correspond to more frequent direct collapse transitions relative to recovery, accelerating the loss of system functionality over time.
\FloatBarrier
This parametrization also provides direct insight into the evolution of the system reliability. For fixed $\mu_{10}  = 1$ and $\varepsilon = 2$, Figure~\ref{fig:toy-beta2}(a) shows the failure probability $ F_{\mathcal{T}}(T_{\text{hor}}) $ as a function of $\alpha$ and $\gamma$. Increasing $\gamma$ raises the rate of direct transitions to collapse, which increases the likelihood of failure within the prescribed time horizon. The effect is further amplified as $\alpha$ increases, since damage accumulates faster relative to recovery. Figure~\ref{fig:toy-beta2}(b) reports the corresponding reliability index $\beta$ computed as described in Eq.~\eqref{eq:beta}. As expected, $\beta$ decreases when either collapse transitions become more frequent or recovery becomes less effective.
Reconsidering here the previously discussed value of $\lambda_{02}$, which is directly tied to seismic design prescriptions, collapse is commonly associated in structural codes with a reference return period, for example $T_{\text{return}} = 475$ years. This corresponds to an equivalent collapse rate of approximately $\lambda_{02} \simeq 2 \times 10^{-3} \, [\text{year}]^{-1}$, identified by the vertical line in Figure~\ref{fig:toy-beta2}(b). Intuitively, moving to the right of this line corresponds to shorter return periods and higher collapse rates $\lambda_{02}$, resulting in a rapid loss of reliability.
For a fixed collapse rate, variations in $\alpha$ isolate the effect of recovery efficiency relative to damage accumulation. Larger values of $\alpha$ indicate that damage accumulates faster than recovery actions can mitigate it, resulting in a marked reduction of the reliability index $ \beta$, even when collapse hazard remains unchanged. 

\begin{figure}
    \centering
    \includegraphics[width=1.0\linewidth]{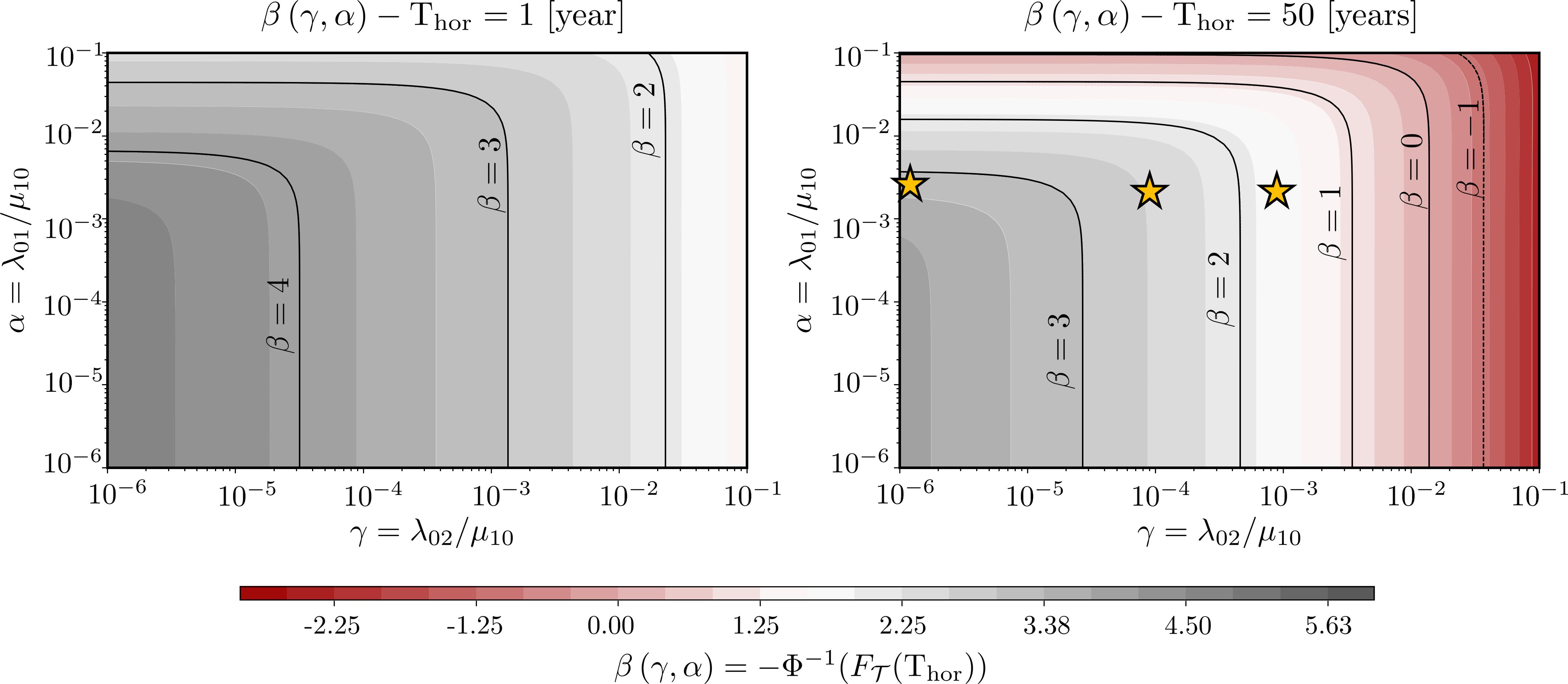}
    \caption{Reliability index $\beta(\gamma,\alpha)$ for $T_{\text{hor}} = 1\,\text{year}$ (left) and $T_{\text{hor}} = 50\,\text{years}$ (right), with $\mu_{10} = 1\,[\text{year}]^{-1}$ and $\varepsilon = \lambda_{12}/\lambda_{01}$ fixed. The color scale ranges from low (red) to high (grey) reliability. Contour lines mark integer values of $\beta$. Three reference points at $\beta = 1,2,3$ (gold stars) are identified on the 50-year map and used as anchors in Figure~\ref{fig:beta-rho-map}.}
    \label{fig:beta-map}
\end{figure}

\begin{figure}
    \centering
    \includegraphics[width=1.0\linewidth]{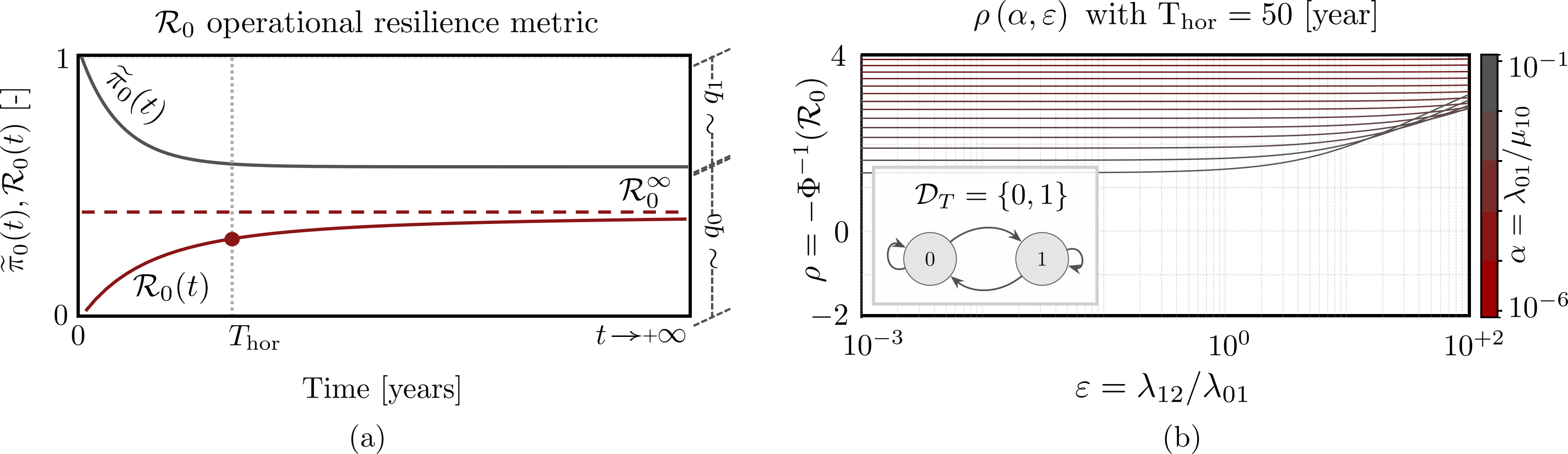}
    \caption{(a) Illustration of the time evolution of the conditional state probability $\tilde{\pi}_0(t)$ and the resilience metrics $\mathcal{R}_0(t)$. The dashed vertical line marks $T_{\text{hor}}$. As $t \to \infty$, the conditioned distribution converges to the quasi-stationary distribution with weights $q_0$ and $q_1$ on the undamaged and damaged states, respectively (Eq.~\eqref{eq:tilde_pi_qsd_main}). (b) Resilience index $\rho(\alpha,\varepsilon)$ at $T_{\text{hor}} = 50\,\text{year}$, as a function of $\varepsilon$, for increasing $\alpha$ (color scale, dark to red). }
    \label{fig:toy-R0-rho}
\end{figure}

\begin{figure}
    \centering
    \includegraphics[width=0.8\linewidth]{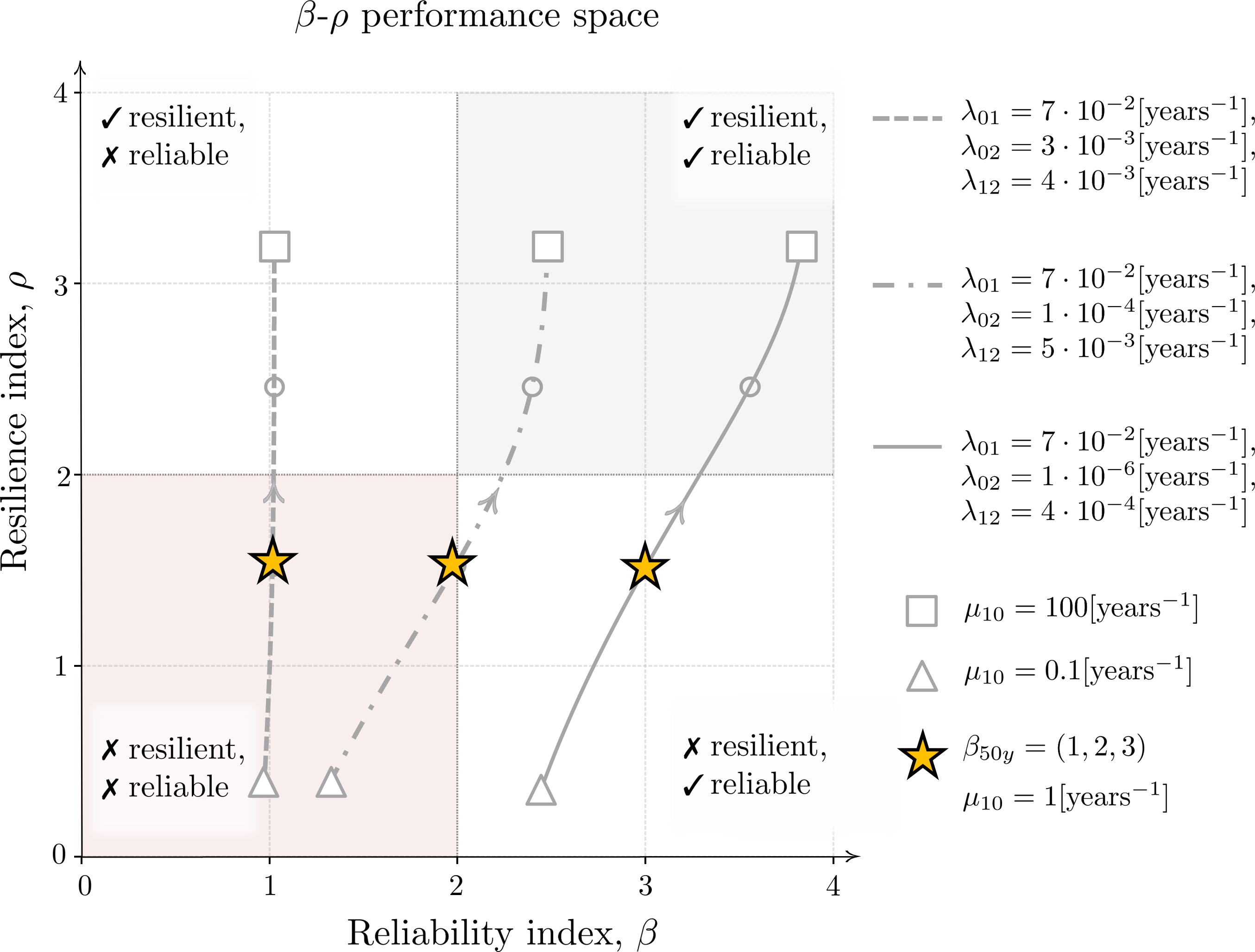}
    \caption{Joint $(\beta,\rho)$ performance space. Three structural configurations with fixed $(\lambda_{01},\lambda_{02},\lambda_{12})$ rates and varying recovery rate $\mu_{10}$ (triangles: $\mu_{10}=0.1\,[\text{year}]^{-1}$; circles: $\mu_{10}=1\,[\text{year}]^{-1}$; squares: $\mu_{10}=100\,[\text{year}]^{-1}$), evaluated at $T_{\text{hor}}=50\,\text{years}$. Arrows indicate the direction of increasing $\mu_{10}$. Gold stars mark the three reference configurations at $\beta=1,2,3$ with $\mu_{10}=1\,[\text{year}]^{-1}$ identified in Figure~\ref{fig:beta-map}.} 
    \label{fig:beta-rho-map}
\end{figure}

Moreover, Figure~\ref{fig:beta-map} shows the reliability index $\beta(\gamma,\alpha)$ for two time horizons: $T_{\text{hor}} = 1\,\text{year}$ (left) and $T_{\text{hor}} = 50\,\text{years}$ (right), with $\mu_{10} = 1\,[\text{year}]^{-1}$ fixed throughout. The contour structure reveals a clear asymmetry between the two parameters. The direct-collapse rate $\gamma$ drives a sharp horizontal transition: even moderate increases in $\gamma$ push the system into negative $\beta$ territory at the 50-year horizon, reflecting the compounding effect of rare but irreversible events over long exposure periods. By contrast, $\alpha$, which governs the competition between damage accumulation and recovery, produces smoother, more gradual variations in $\beta$. This is physically expected: transitions through the damaged state $ds_1$ allow recovery to partially counteract damage accumulation, whereas direct collapse from $ds_0$ is immediately absorbed with no possibility of return.
On the 50-year contour map, three representative points are marked at $\beta = 1,2,3$, identified at fixed $\alpha$ and increasing $\gamma$. These three configurations are carried forward to Figure~\ref{fig:beta-rho-map}, where they serve as reference anchors in the joint $(\beta,\rho)$ performance space.
Figure~\ref{fig:toy-R0-rho}(a) illustrates the time evolution of the conditional state probability $\tilde{\pi}_0(t)$, defined in Eq.~\eqref{eq:tilde_pi_qsd_main} as the probability of occupying the undamaged state $ds_0$ given that no collapse has occurred by time $t$. Starting from $\tilde{\pi}_0(0)=1$, this quantity decays monotonically and converges, as $t\to\infty$, to the quasi-stationary distribution (QSD) weight $q_0=\nu_0$, with its complement $q_1=\nu_1$ indicated on the vertical axis. Superimposed on the same axes, the resilience metric $\mathcal{R}_0(t)$ of Eq.~\eqref{eq:conditional_mean_occupation_fraction} is shown as a function of time. As $t$ grows, $\mathcal{R}_0(t)$ converges to the spectral approximation $\mathcal{R}_0^\infty = 1 - w_0\,q_0$ derived in Eq.~\eqref{eq:R0_limit_complement_main}, a value determined solely by the spectral structure of $\bm Q_T$ and independent of the initial condition. 

Figure~\ref{fig:toy-R0-rho}(b) shows the resilience index $\rho = -\Phi^{-1}(\mathcal{R}_0)$ (Eq.~\eqref{eq:rho}) as a function of $\varepsilon = \lambda_{12}/\lambda_{01}$ for increasing $\alpha$, at $T_{\text{hor}} = 50\,\text{year}$. For small $\varepsilon$, the curves are nearly parallel: collapse from the damaged state is rare, and resilience is governed primarily by the damage--recovery competition encoded in $\alpha$. As $\varepsilon$ grows, all curves bend sharply downward and converge. The engineering interpretation is direct: once $\lambda_{12} = \varepsilon\,\alpha\,\mu_{10}$ becomes comparable to $\mu_{10}$, the system is more likely to collapse than to recover from $ds_1$, regardless of how frequently damage is repaired. This is the regime of brittle or capacity-limited structures --- the intervention window between damage onset and collapse is simply too short for repair actions to be effective.

The joint $(\beta,\rho)$ performance space brings together reliability and resilience into a single diagram. Figure~\ref{fig:beta-rho-map} shows three curves, each corresponding to a fixed set of absolute seismic rates $(\lambda_{01}, \lambda_{02}, \lambda_{12})$, traced as the recovery rate $\mu_{10}$ increases from $0.1\,[\text{year}]^{-1}$ (slow repair, triangles) to $100\,[\text{year}]^{-1}$ (near-instantaneous repair, squares), spanning timescales from years down to days. The three parameter sets are chosen to represent qualitatively distinct structural configurations, differing primarily in $\lambda_{02}$ and $\lambda_{12}$, i.e., in the rates governing irreversible collapse.

The trajectories in the $(\beta,\rho)$ space reveal a fundamental decoupling between the two indices. Increasing $\mu_{10}$ moves each system predominantly along the $\rho$ axis, with only a modest gain in $\beta$: faster recovery reduces the time spent in damaged states, improving resilience, but leaves the seismic collapse rate, encoded in $\lambda_{02}$ and $\lambda_{12}$, unchanged. This decoupling is a structural property of the system: reliability is governed by the damage-to-collapse rates, while resilience is governed by the competition between damage and recovery.

The three reference points from Figure~\ref{fig:beta-map} at $\beta = 1,2,3$ with $\mu_{10} = 1\,[\text{year}]^{-1}$ are superimposed as gold stars. Their positions illustrate how the same recovery rate produces very different resilience outcomes depending on the configuration. A system with $\beta = 1$ --- low reliability --- gains almost no $\beta$ improvement as $\mu_{10}$ increases, and moves nearly vertically toward higher $\rho$: recovery alone cannot compensate for a high collapse rate. A system with $\beta = 2$ benefits more substantially from faster repair, achieving meaningful joint improvement in both indices as it moves toward the upper-right quadrant. A system already at $\beta = 3$ --- high reliability --- is so rarely damaged that even moderate recovery rates suffice to push $\rho$ to large values; further increases in $\mu_{10}$ yield diminishing returns on $\beta$ but continued gains in resilience.

These trajectories delineate the four performance quadrants visible in the figure. The upper-right region ($\beta$ high, $\rho$ high) represents the target for critical infrastructure: collapse is unlikely and damage is repaired quickly. The lower-right region ($\beta$ high, $\rho$ low) corresponds to structures with adequate collapse safety but slow or absent repair --- a condition common in monumental or heritage buildings where structural integrity is preserved but restoration is constrained by non-technical factors. The upper-left region ($\beta$ low, $\rho$ high) identifies operationally robust but structurally fragile systems, such as lightly constructed buildings in moderate-hazard zones with rapid repair protocols. The lower-left region ($\beta$ low, $\rho$ low) is the most critical: the system collapses frequently and spends most of its life in a damaged condition.

\newpage

\section{Application to two archetypal industrial systems}
\label{sec:case-study}
\subsection{Time-invariant seismic hazard} \vspace{-2mm}
We adopt a time-invariant seismic hazard model, consistent with (Hyp~\ref{hyp:indep_event}) and (Hyp~\ref{hyp:stat}), which exclude aftershock/foreshock sequences and temporal clustering. This allows a time-homogeneous CTMC formulation; introducing time dependence would require a non-homogeneous CTMC, outside the scope of this work.

The hazard scenario refers to L'Aquila (central Italy), with a 10\% probability of exceedance in 50 years ($T_{\text{return}} \approx 475$ years). Annual exceedance rates and their 5$^{\text{th}}$--95$^{\text{th}}$ percentile bounds are sourced from the ESHM \cite{bib:danciu2021eshm20} via the EFEHR portal.\footnote{\url{https://hazard.efehr.org/en/hazard-data-access/hazard-curves/}} Geographic location and hazard curves are reported in Appendix~\ref{app:spif}.

\subsection{Generator rate matrix $\boldmath Q =\boldmath Q_{\text{dam}} + \boldmath Q_{\text{rec}}$}
\begin{figure}[!ht]
    \centering
    \includegraphics[width=0.40\linewidth]{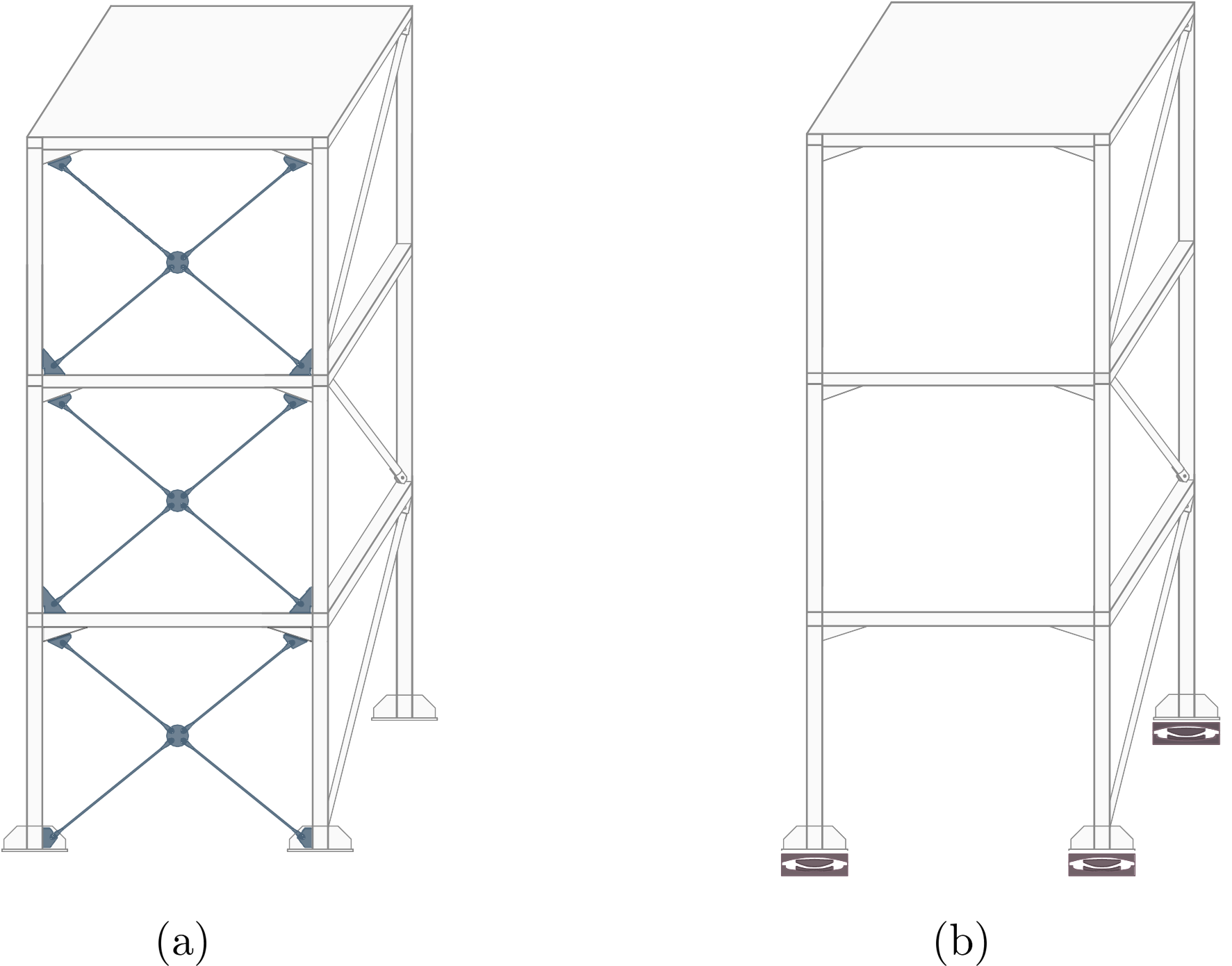}
    \caption{Simplified sketches of the case study structures: (a) steel braced frame (BF); (b) steel base isolated (BI) frame equipped with a triple friction pendulum system.}
    \label{fig:mockups}
\end{figure}

We consider two structural models based on industrial-related case studies, sketched in Figure~\ref{fig:mockups}. The first is a full-scale, three-story steel braced frame (BF) designed for a large-scale experimental campaign conducted at EUCENTRE (Pavia, Italy), comprehensively described both in configuration and numerical modeling in \cite{bib:nardin2022BF, bib:nardin2022thesis}. The second structure represents a base-isolated (BI) version of the frame without bracing system, equipped with a triple friction pendulum isolator, reflecting standard engineering practice for high demanding isolation solutions \cite{bib:maurerSIPadaptive}. Further design and modeling specifics for this case are reported in \cite{bib:butenweg2020sera}.

\subsubsection{Submatrix $\boldmath Q_{\text{dam}}$: seismic state-dependent fragilities} \label{sec:case-studies}
\begin{figure}[!ht]
    \centering
    \includegraphics[width=1.\linewidth]{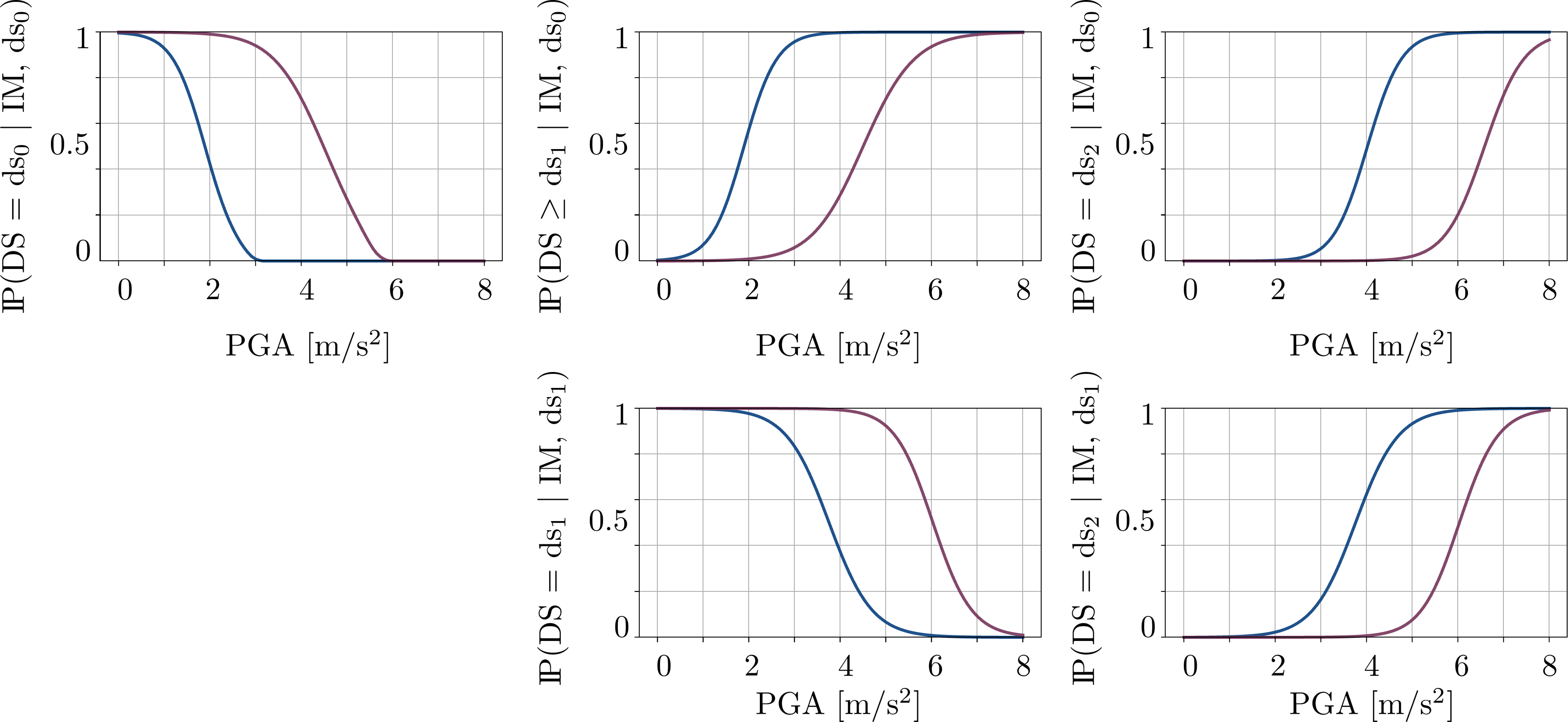}    
    \caption{State-dependent fragility functions for the two case-study systems, adapted from Nardin \textit{et al.} \cite{bib:nardin2025ress}. Blue curves correspond to the steel BF, and purple curves to the BI configuration.}
    \label{fig:state-dependent-fragilities}
\end{figure}

\vspace{-2mm}
We adopt the median state-dependent fragility curves shown in Figure~\ref{fig:state-dependent-fragilities}, derived from the seismic performance of the two structural systems. For the braced frame (blue curves), the fragility functions were computed by the authors following the procedure described in \cite{bib:nardin2025ress}.

For the base-isolated configuration (purple curves), no dedicated fragility analysis was performed. Instead, the braced frame curves are manually shifted to higher intensity levels, reflecting the well-established engineering understanding that base-isolated systems outperform their fixed-base counterparts due to period lengthening and increased energy dissipation, which collectively reduce seismic demand and raise median capacities \cite{bib:thippa2024, bib:Kitayama2023downtime}. This shift is therefore a qualitative, judgment-based adjustment, not a quantitative outcome of structural analysis, and is intended solely for illustrative purposes within the CTMC framework.

This pragmatic choice allows us to focus on the core methodological pipeline and to demonstrate how state-dependent fragilities can be seamlessly embedded within the proposed framework and provided code.
\FloatBarrier
Following the original literature in~\cite{bib:butenweg2020sera}, the damage classification comprises three discrete states: $ds_0$ (undamaged), $ds_1$ (slightly damaged, equivalent to the serviceability limit state), and $ds_2$ (severely damaged, equivalent to the ultimate limit state).

For the braced frame configuration, drift-based limit state thresholds are defined according to FEMA guidelines~\cite{bib:fema356} and are summarized in \tabref{tab:edp_threshold}. In contrast, for the base-isolated configuration, the ASCE 7-16 standard~\cite{bib:asce17chapter17} specifies the maximum allowable story drift above the isolation system. Additional thresholds for the isolator maximum slips are derived from technical documentation of the specific device employed~\cite{bib:maurerSIPadaptive}, as reported in \tabref{tab:edp_threshold}.

\begin{table}[!ht]
  \centering
  \caption{EDP and damage state thresholds for the two case studies.}
  \scalebox{0.80}{
  \begin{tabular}{l l l l l}
    \toprule
    \makecell[c]{\textbf{Structure}\\\textbf{configuration}} &
    \makecell[c]{\textbf{EDP}\\\textbf{threshold}} &
    \makecell[c]{\textbf{ds\textsubscript{0}}} &
    \makecell[c]{\textbf{ds\textsubscript{1}}} &
    \makecell[c]{\textbf{ds\textsubscript{2}}} \\
    \midrule
    Brace frame &
    max. drift \textcolor{white}{or slip} &
    $< 0.5\%\, h_{\text{sx}}$ &
    $0.5\%\text{–}1.5\%\, h_{\text{sx}}$ &
    $> 2\%\, h_{\text{sx}}$ \\
    \midrule
    \multirow{2}{*}{Base isolated} &
    max. isolator slip &
    $<100$ mm &
    $100\text{–}200$ mm &
    $>200$ mm \\
    & \makecell[c]{max. drift \\@ superstruct.} &
    \multicolumn{3}{c}{\makecell[c]{$<1.5\% \,h_{\text{sx}}$ with $h_{\text{sx}}$ storey heights}} \\
    \bottomrule
  \end{tabular}}
  \label{tab:edp_threshold}
\end{table}

\FloatBarrier

\subsubsection{Submatrix $\boldmath Q_{\text{rec}}$: seismic recovery processes}\label{sub:recovery} \vspace{-2mm}
The recovery process in our model is represented as a single state transition, that is from the damaged state, $ds_1$, back to the fully functional state, $ds_0$, governed by a constant repair rate, $\mu_{10}$. Because direct empirical measurements of such rates are scarce, we investigate a plausible range of values informed by established post-disaster recovery guidelines (e.g., the FEMA P-58~\cite{bib:fema2018p58_1}, the HAZUS taxonomy~\cite{bib:fema2020hazus}, etc.), supplemented by engineering judgment. 

Both HAZUS and FEMA P-58 provide repair-time estimates, either empirically derived or judgment-based, that aggregate risk information for large building inventories. These are typically organized into libraries of median repair times or downtimes, which have been incorporated into computational tools such as PELICUN~\cite{bib:Zsarnoczay2024PBE}\footnote{Freely available at \url{https://github.com/OpenPBEE/PBEE-Recovery}.} and the integrated ATC-138 inventory~\cite{bib:ATC138_2021} PACT software~\cite{bib:fema2018p58_2}.

For this study, representative median repair times were compiled as summarized in Table~\ref{tab:repair_times}. For steel braced frames, values were adopted from HAZUS and ATC guidance for industrial facilities, which provide hazard-dependent median repair times. Specifically, median durations corresponding to low- and moderate-to-severe-damage cases were considered in order to explore the influence of variability in recovery timing. No standardized data currently exist for base-isolated structures; however, their recovery times are generally expected to be shorter because properly designed seismic isolation reduces both structural and nonstructural damage \cite{bib:asce17chapter17}. This expectation is supported by recent studies \cite{bib:TerzicMahin2017, bib:Kitayama2023downtime}, which report that base-isolated steel buildings typically require only a few hours to several days to recover from minor events, and about four to six weeks ($\approx$50 days) following strong shaking. 

The values in Table~\ref{tab:repair_times} define the range of recovery durations explored in our simulations, categorized into short-term and long-term recovery scenarios to reflect varying recovery efficiencies and damage severities. To integrate these durations into the CTMC framework, we treat the recovery process from state $j$ to state $i$ as a stochastic process with an exponentially distributed holding time. Under this assumption, the constant repair rate $\mu_{ij}$ is uniquely determined by the median recovery time $t_{\mathrm{med}}$. Therefore, the transition rate is computed as:
\begin{equation}
\mu_{10} = \frac{\ln(2)}{t_{\mathrm{med}}} \cdot 365
\end{equation}
where $t_{\mathrm{med}}$ is expressed in days and the factor 365 annualizes the rate to maintain consistency with the seismic hazard units (events/year). 

\sisetup{
  table-number-alignment = center
}
\setcounter{table}{2} 
\begin{table}[h!]
\centering
\caption{Median recovery times $t_{\mathrm{med}}$ and repair rates $\mu_{10}$ for the BF and BI systems.}
\label{tab:repair_times}
\scalebox{0.85}{
    \begin{tabular}{
      l
      S[table-format=3.0]
      S[table-format=3.2]
      c
      c
    }
    \toprule
    \multirow{2}{*}{\textbf{Configuration}} &
    \multicolumn{2}{c}{\textbf{Short-term recovery}} &
    \multicolumn{2}{c}{\textbf{Long-term recovery}} \\
    \cmidrule(lr){2-3} \cmidrule(lr){4-5}
     & {$t_{\mathrm{med}}$ [days]} & {$\mu_{10}$ [year$^{-1}$]} 
     & {$t_{\mathrm{med}}$ [days]} & {$\mu_{10}$ [year$^{-1}$]} \\
    \midrule
    Steel braced frame (BF)   & 30  & 8.43   & 120--365 & 2.10--0.69 \\
    Base-isolated system (BI) & 1   & 252.99 & 30--50   & 8.42--5.05 \\
    \bottomrule
    \end{tabular}}
\end{table}


\subsection{CTMC analysis for system reliability and resilience}\label{sec:discussion}
With all the case-study parameters defined, we now evaluate the performance of the two structural systems under repeated seismic hazard and recovery processes. The analysis unfolds along three complementary dimensions: (i) long-term probability evolution across damage and collapse states; (ii) reliability analysis via the transient generator $\bm Q_T$; and (iii) resilience analysis through conditional state occupancy and spectral properties of $\bm Q_T$.

\subsubsection{Long-term probability dynamics}
\begin{figure}[!ht]
    \centering
    \includegraphics[width=0.99\linewidth]{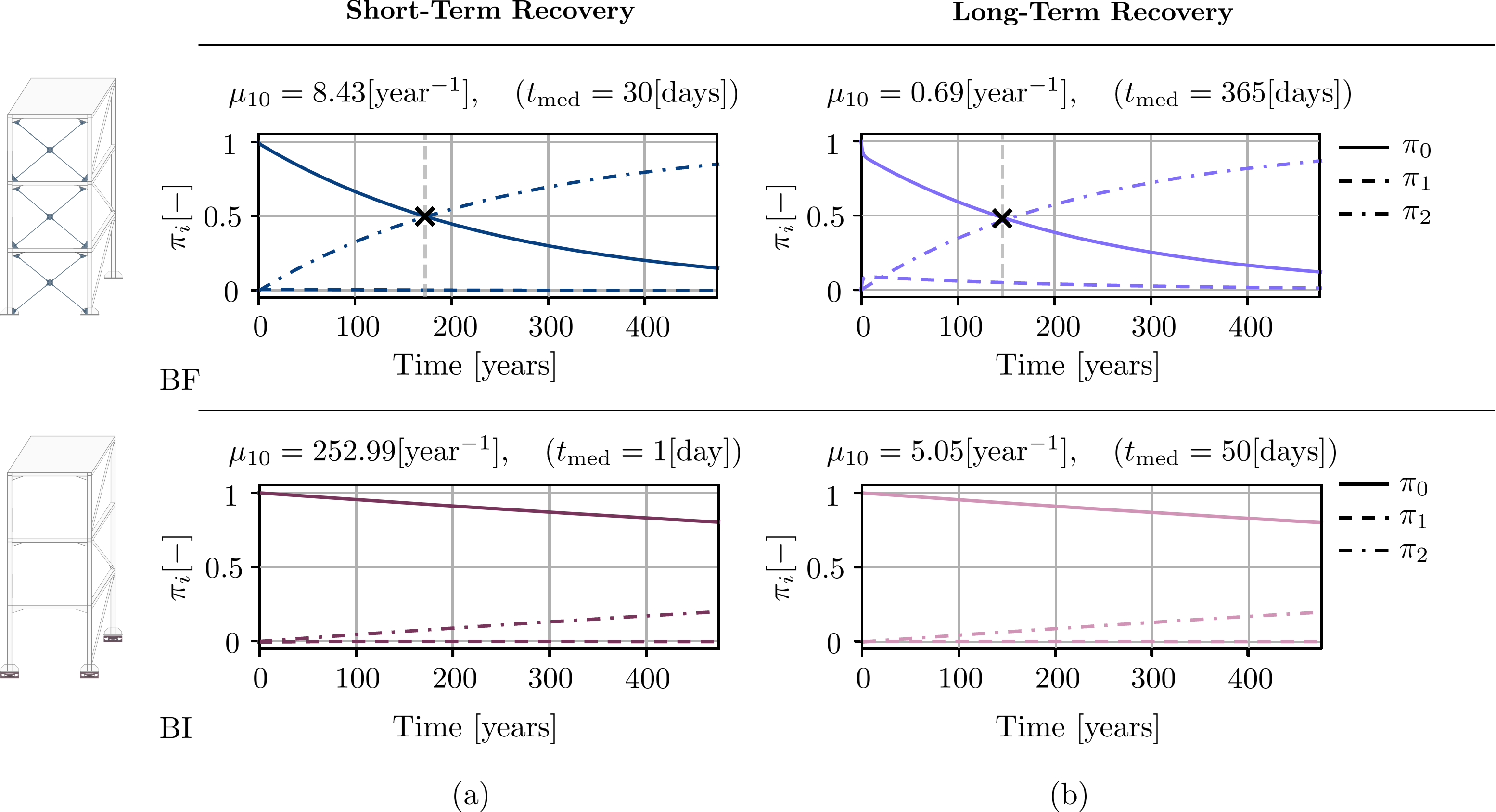}
    \caption{Time evolution of state probabilities $\pi_i(t)$ for the braced frame (BF, top) and base-isolated (BI, bottom) systems under short- and long-term recovery scenarios. The vertical dashed line marks the approximate time when $\pi_0=\pi_2$, highlighting the delayed condition in the isolated configuration.}
    \label{fig:full-distribution-mockups}
\end{figure}

Figure~\ref{fig:full-distribution-mockups} illustrates the time evolution of $\pi_i(t)=\mathbb{P}\!\left(DS(t)=ds_i\right)$ up to $T_{\mathrm{return}} = 475\,[\text{years}]$. The upper row refers to the braced frame (BF) and the lower row to the base-isolated (BI) system; columns (a) and (b) correspond to short- and long-term recovery scenarios (Table~\ref{tab:recovery-distribution}).
 
A striking contrast emerges between the two systems. For the BF, the crossover $\pi_0(t)=\pi_2(t)$ occurs at roughly 150 years, after which the collapse state becomes the dominant occupancy. For the BI system, by contrast, no such crossover is observed within the return period $T_{\mathrm{return}}=475$ years: $\pi_0(t)$ remains well above $\pi_2(t)$ throughout, reflecting the substantially lower damage accumulation rates achieved by the isolation system. This qualitative difference underscores how base isolation not only delays but effectively suppresses the transition to a collapse-dominated regime over a standard structural life cycle.
 
A particularly notable feature of Figure~\ref{fig:full-distribution-mockups}, further confirmed by the state-probability distributions reported in Table~\ref{tab:recovery-distribution}, is the markedly different sensitivity to the recovery rate $\mu_{10}$ between the two configurations. For the BI system, the curves corresponding to short- and long-term recovery scenarios are  indistinguishable at all time scales, and the state-probability vectors converge to essentially identical values regardless of the recovery scenario. This insensitivity reflects the very short recovery timescales of the BI system, $\mu_{10}=(5,\,253)\,[\text{year}^{-1}]$, corresponding to recovery times from a single day to a few months, which are orders of magnitude shorter than the mean inter-arrival time of damaging events ($\sim 10^3\,[\text{years}]$): damaged states are repaired almost immediately, so $\mu_{10}$ has negligible influence on the long-term drift toward $ds_2$.

For the BF, by contrast, the state-vector distribution is influenced by recovery  rate $\mu_{10}$. At $t=50\,[\text{years}]$, the short- and long-term recovery scenarios yield state distributions of $(0.81,\,0.01,\,0.18)$ and $(0.73,\,0.08,\,0.19)$, respectively, and the gap remains still visible at $t=100\,[\text{years}]$ even if reduced. This residual sensitivity arises because the BF recovery timescales are less cleanly separated from the damage arrival rates than in the BI case, so the competition between repair and re-damage is non-negligible. 

\begin{table}[!ht]
  \centering
  \caption{State-probability distributions at $t=50$ years and $t=100$ years for short- ($t_{\text{med}}=30\ \text{days}$ for the BF, and $t_{\text{med}}=1\ \text{day}$ for the BI, respectively) and long- ($t_{\text{med}}=365\ \text{days}$ for the BF, and $t_{\text{med}}=50\ \text{days}$ for the BI, respectively) term recovery scenarios.}
  \label{tab:recovery-distribution}
  \scalebox{0.85}{
  \begin{tabular}{rccrcc}
    \cmidrule(lr){2-6}
          & \multicolumn{5}{c}{\textbf{Recovery rate (time)}} \\
    \cmidrule(lr){2-6}
          & \textbf{Short-term} & \textbf{Long-term} & & \textbf{Short-term} & \textbf{Long-term} \\
    \midrule
    \textbf{Braced frame}  & $(0.81,\,0.01,\,0.18)$ & $(0.73,\,0.08,\,0.19)$ & & $(0.67,\,0.01,\,0.33)$ & $(0.59,\,0.06,\,0.35)$ \\
    \cmidrule(lr){1-3}\cmidrule(lr){5-6}
    \textbf{Base-isolated}  & $(0.99,\,0.00,\,0.01)$ & $(0.98,\,0.00,\,0.02)$ & & $(0.95,\,0.00,\,0.05)$ & $(0.95,\,0.00,\,0.05)$ \\
    \midrule
          & \multicolumn{2}{c}{$t=50\ \text{years}$} & & \multicolumn{2}{c}{$t=100\ \text{years}$} \\
  \end{tabular}}
\end{table}

\subsubsection{Structural reliability analysis under seismic hazard}\label{sub:reliability-discussion}
\begin{figure}[!ht]
    \centering
    \includegraphics[width=1.0\linewidth]{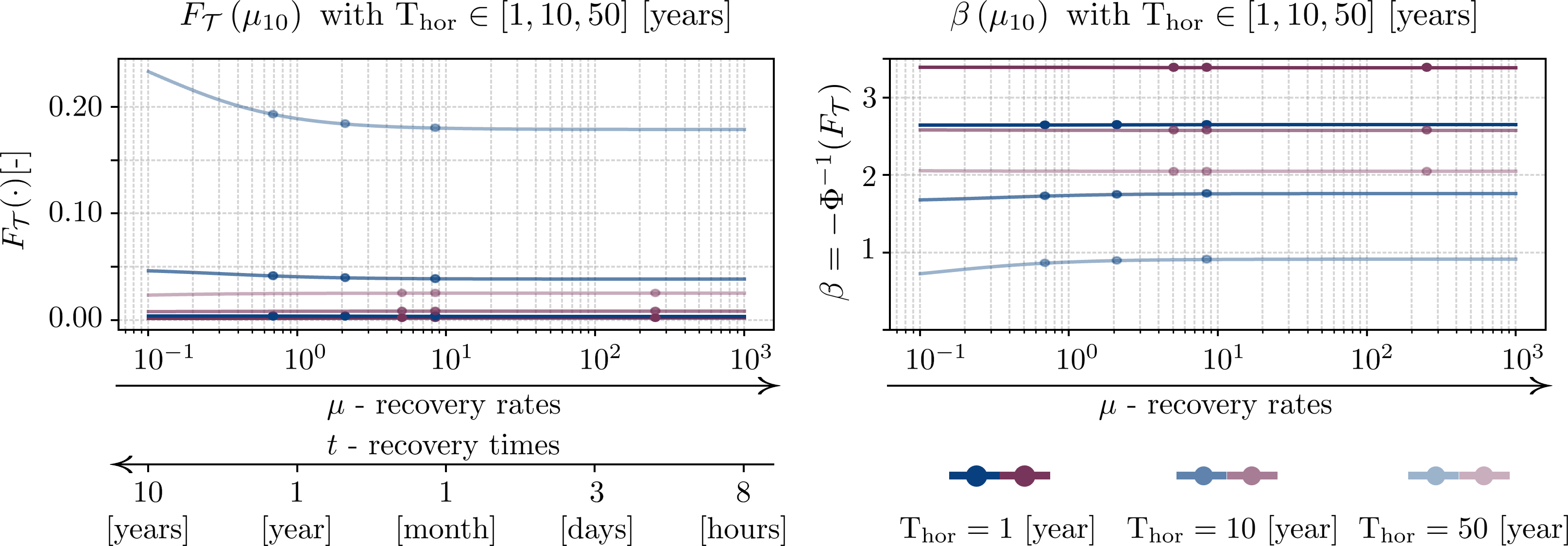} 
    \caption{$F_{\mathcal{T}}(\cdot)$ (left) and corresponding $\beta$ (right) as functions of the recovery rate $\mu_{10}$ and the reference time horizon $T_{\text{hor}}$. Results for the BF (on the left) and for the BI (on the right). Color intensity denotes the time horizon, corresponding to 1, 10, and 50 years. Filled markers indicate the median recovery rates reported in Table~\ref{tab:repair_times}.}
    \label{fig:ftau-beta-industrial}
\end{figure}

We now examine the failure probability $F_{\mathcal{T}}(\cdot)$ defined in Eq.~\eqref{eq:failure_F_tau} and the reliability index $\beta_t = -\Phi^{-1}(F_{\mathcal{T}}(t))$ of Eq.~\eqref{eq:beta}, sweeping the recovery rate $\mu_{10}$ over the interval from $10^{-1}$ up to $10^{3}\,[\text{year}^{-1}]$ while keeping all seismic rates fixed. Figure~\ref{fig:ftau-beta-industrial} summarizes the results.
 
The BF consistently exhibits higher failure probabilities than the BI system, reflecting greater damage accumulation under repeated loading. For both configurations, $\beta_t$ shows limited sensitivity to $\mu_{10}$: as anticipated from the timescale argument above and from the plateau region in Figure~\ref{fig:toy-beta2} of Section~\ref{sec:toy-example}, long-term reliability is governed by the seismic undamage/damage-to-collapse rates $\lambda_{02}$ and $\lambda_{12}$ rather than by post-event repair speed.
A result of particular significance is that, at the 50-year horizon, the BF yields $\beta_{50} \approx 1$. This low value should be interpreted in context: the braced frame was designed for a large-scale shake-table experimental campaign, whose primary objective was the study of the coupling between structural response and non-structural industrial components~\cite{bib:nardin2022BF, bib:nardin2022thesis}, not the 
optimization of system reliability. The marked gap in $\beta_{50}$ between the BF and BI configurations therefore reflects partly the experimental origin of the BF design, and should not be taken as a general quantitative benchmark between fixed-base and base-isolated archetypes.

\subsubsection{Spectral and seismic resilience analysis} \label{sub:spectral-discussion}
We now quantify resilience through the conditional occupation fraction $\mathcal{Q}_0(t)$ of Eq.~\eqref{eq:Q0_def} and its complement $\mathcal{R}_0(t)$ of Eq.~\eqref{eq:conditional_mean_occupation_fraction}, mapped to the standardized resilience index $\rho_t = -\Phi^{-1}(\mathcal{R}_0(t))$ via Eq.~\eqref{eq:rho}. The same quantities are also obtained from the spectral approximation $\mathcal{R}_0 \approx 1 - w_0\,q_0$ of Eq.~\eqref{eq:R0_limit_complement_main}, valid when the system operates in a quasi-stationary regime before absorption.
 
The spectral approximation is applicable when the characteristic decay time $1/|\sigma_1|$ substantially
exceeds $T_{\mathrm{hor}}$, and the spectral gap is large enough that higher-order modes have
decayed. Table~\ref{tab:resilience-50years} reports the eigenvalues $\sigma_1$, $\sigma_2$, and the spectral
ratio $|\sigma_2/\sigma_1|$ for all configurations given 50 years. For the BF, $|\sigma_2/\sigma_1|\sim
10^{2}$--$10^{3}$; for the BI, $|\sigma_2/\sigma_1|\sim 10^{3}$--$10^{5}$. In all cases $T_{\mathrm{hor}}\ll
1/|\sigma_1|$, confirming that the quasi-stationary approximation is accurate with worst-case discrepancies
$\Delta\mathcal{R}_0\sim 10^{-2}$.


\setlength{\tabcolsep}{1.5pt}
\begin{table}[!ht]
  \centering
  \caption{\raggedright Resilience metrics evaluated at $T_{\text{hor}} = 50$ years for the BF and BI systems. The resilience index $\mathcal{R}_0$ and standardized resilience $\rho$ are reported for short- and long-term recovery scenarios. Results are obtained from both the exact conditional formulation (Eq.\eqref{eq:conditional_mean_occupation_fraction}) and the spectral approximation (Eq.\eqref{eq:R0_limit_complement_main}).}
  \scalebox{0.75}{
    \begin{tabular}{p{1.5em}cccp{0.5em}cccp{1.5em}cccp{0.5em}ccc}

          & \multicolumn{3}{c}{\textbf{ Short-term}} 
          & & \multicolumn{3}{c}{\textbf{ Long-term}} 
          & & \multicolumn{3}{c}{\textbf{ Short-term}} 
          & & \multicolumn{3}{c}{\textbf{ Long-term}} \\

          & \multicolumn{3}{c}{\boldmath{}\textbf{$\mu_{10}=8.43\;[\text{year}^{-1}]$}\unboldmath{}} 
          && \multicolumn{3}{c}{\boldmath{}\textbf{$\mu_{10}=0.69\;[\text{year}^{-1}]$}\unboldmath{}}

          && \multicolumn{3}{c}{\boldmath{}\textbf{$\mu_{10}=253.00\;[\text{year}^{-1}]$}\unboldmath{}}
          && \multicolumn{3}{c}{\boldmath{}\textbf{$\mu_{10}=5.06\;[\text{year}^{-1}]$}\unboldmath{}} \\

    \cmidrule{1-8}\cmidrule{10-16}

    \multicolumn{1}{l}{\boldmath{}\textbf{Spectral }} 
      & \multicolumn{3}{c}{$\sigma_1=0.0035,\;\sigma_2=8.4933$}
      && \multicolumn{3}{c}{$\sigma_1=0.0037,\;\sigma_2=0.7530$}

      && \multicolumn{3}{c}{$\sigma_1=0.0004,\;\sigma_2=253.0035$}
      && \multicolumn{3}{c}{$\sigma_1=0.0004,\;\sigma_2=5.0648$} \\

    \multicolumn{1}{l}{\boldmath{}\textbf{properties }} 
      & \multicolumn{3}{c}{$|\sigma_2/\sigma_1| \sim 10^{3},\;\beta=0.998$}
      && \multicolumn{3}{c}{$|\sigma_2/\sigma_1| \sim 10^{2},\;\beta=0.960$}

      && \multicolumn{3}{c}{$|\sigma_2/\sigma_1| \sim 10^{5},\;\beta=2.082$}
      && \multicolumn{3}{c}{$|\sigma_2/\sigma_1| \sim 10^{4},\;\beta=2.081$} \\

    \cmidrule{1-8}\cmidrule{10-16}

    \multicolumn{1}{l}{\boldmath{}\textbf{$\mathcal{R}_0$, Eq.\unboldmath{}~\eqref{eq:conditional_mean_occupation_fraction}}}
      & \multicolumn{3}{c}{0.0062}
      && \multicolumn{3}{c}{0.0685}

      && \multicolumn{3}{c}{0.0000}
      && \multicolumn{3}{c}{0.0008} \\

    \multicolumn{1}{l}{\boldmath{}\textbf{$\mathcal{R}_0$, Eq.}\unboldmath{}~\eqref{eq:R0_limit_complement_main}}
      & \multicolumn{3}{c}{0.0062}
      && \multicolumn{3}{c}{0.0703}

      && \multicolumn{3}{c}{0.0000}
      && \multicolumn{3}{c}{0.0008} \\

    \multicolumn{1}{l}{\boldmath{}\textbf{$\Delta \mathcal{R}_0$}\unboldmath{}}
      & \multicolumn{3}{c}{$\sim 10^{-5}$}
      && \multicolumn{3}{c}{$\sim 10^{-3}$}

      && \multicolumn{3}{c}{$\sim 10^{-9}$}
      && \multicolumn{3}{c}{$\sim 10^{-6}$} \\

    \cmidrule{1-8}\cmidrule{10-16}

    \multicolumn{1}{l}{\boldmath{}\textbf{$\rho$, Eq.}\unboldmath{}~\eqref{eq:rho}}
      & \multicolumn{3}{c}{2.50}
      && \multicolumn{3}{c}{1.49}

      && \multicolumn{3}{c}{4.16}
      && \multicolumn{3}{c}{3.15} \\

    \cmidrule{2-8}\cmidrule{10-16}

          & & \multicolumn{6}{c}{BF}
          & & & \multicolumn{6}{c}{BI} \\

    \end{tabular}
  }
  \label{tab:resilience-50years}
\end{table}

\begin{figure}[!ht]
    \centering
    \includegraphics[width=1.0\linewidth]{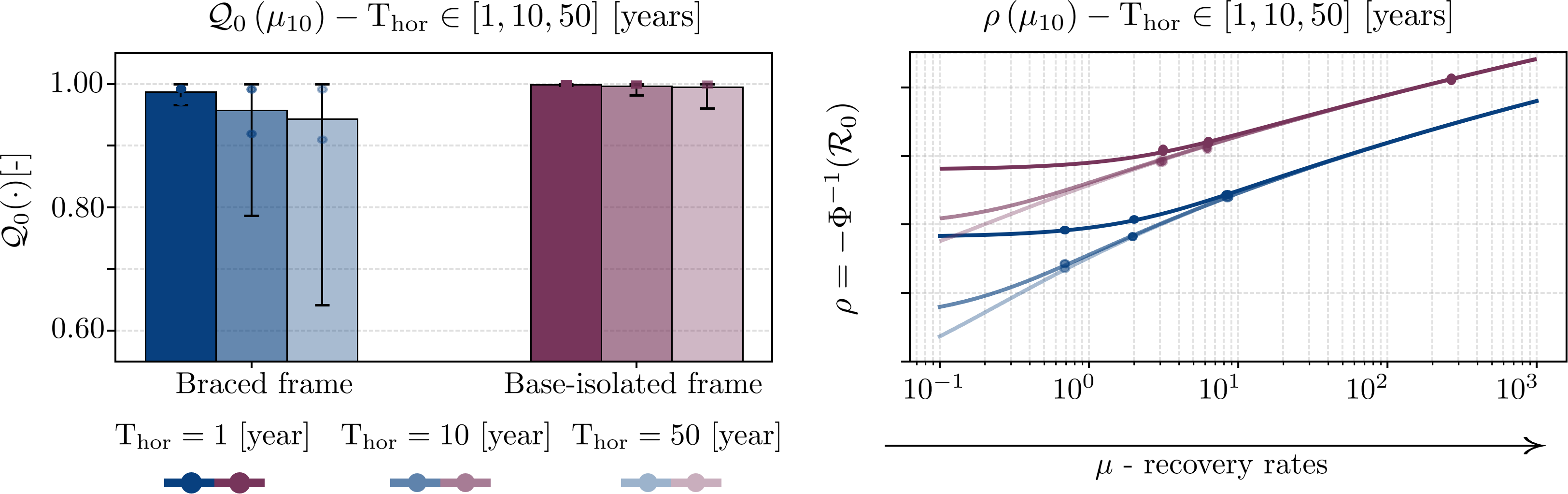} 
    \caption{$\mathcal{Q}_0$ (left) and $\rho$ (right) as functions of the recovery rate $\mu_{10}$ and the reference time horizon $T_{\text{hor}}$. 
    Results are shown for the BF and BI systems. Color intensity denotes the time horizon, corresponding to 1, 10, and 50 years. Filled markers indicate the median recovery rates reported in Table~\ref{tab:repair_times}.}
    \label{fig:resilience-Q0-industrial}
\end{figure}

Figure~\ref{fig:resilience-Q0-industrial} shows $\mathcal{Q}_0$ and $\rho$ as functions of $\mu_{10}$ and $T_{\mathrm{hor}}$. In sharp contrast to the reliability results, both metrics show a pronounced dependence on recovery rate: faster recovery increases the fraction of time spent in $ds_0$ before absorption, directly improving resilience. This confirms that the decoupling between reliability and resilience, first observed in the three-state toy example of Section~\ref{sec:toy-example}, persists in full industrial-scale applications.

The left panel of Figure~\ref{fig:resilience-Q0-industrial} shows, for each 
system and time horizon, the mean conditional occupation fraction $\mathcal{Q}_0$ averaged over the full range of recovery rates $\mu_{10}\in[10^{-1},\,10^{3}]\,[\text{year}^{-1}]$, with error bars spanning the corresponding minimum and maximum values. The markers superimposed on each bar indicate the specific values of $\mathcal{Q}_0$ attained at the reference recovery rates of Table~\ref{tab:repair_times}. 
From the right panel of Figure~\ref{fig:resilience-Q0-industrial} shows the resilience index $\rho$ as a function of $\mu_{10}$ for both systems and different time horizons. 
The BF consistently exhibits lower resilience than the BI system, with the gap widening at longer horizons. For the BI, even the long-recovery scenario ($\mu_{10}\approx 5.06\,\text{year}^{-1}$, $t_{\mathrm{med}}=50\,\text{days}$) yields $\rho\approx 2.56$ at $T_{\mathrm{hor}}=50\,\text{years}$, whereas the short-recovery scenario ($\mu_{10}\approx 253\,\text{year}^{-1}$ $t_{\mathrm{med}}=1\,\text{day}$) reaches $\rho\approx 3.70$ at the same horizon.

These findings confirm the central message of the toy example in Section~\ref{sec:toy-example}: reliability is controlled by the seismic hazard rates and is largely insensitive to $\mu_{10}$, while resilience is strongly governed by the competition between damage and recovery — exactly the decoupling visible in the $(\beta,\rho)$ plane of Figure~\ref{fig:beta-rho-map}.

\begin{figure}[!ht]
    \centering
    \includegraphics[width=0.7\linewidth]{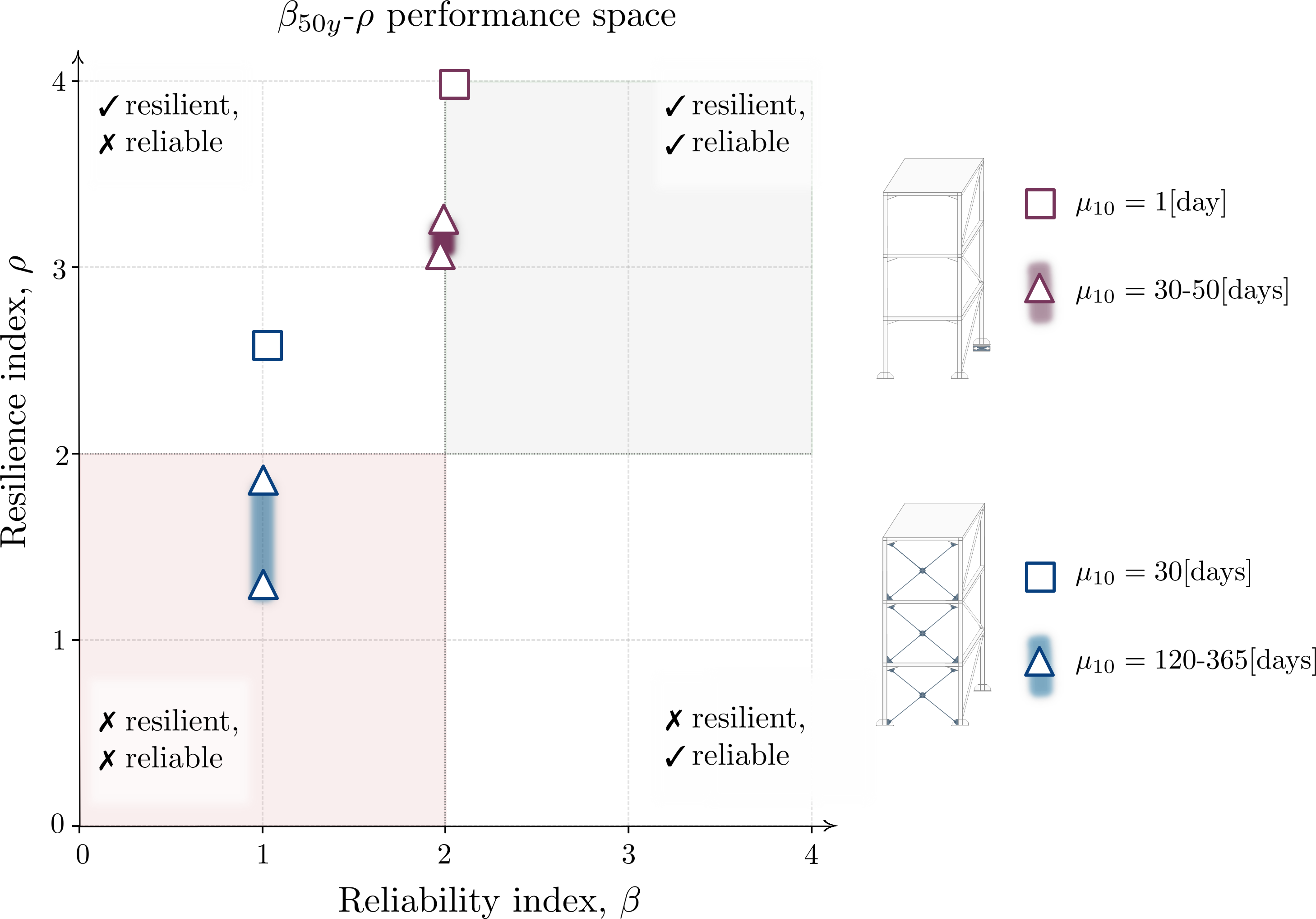} 
    \caption{Joint $(\beta_{50y},\rho)$ performance space for the two industrial case studies. Each configuration is represented by two markers corresponding to the short-term (squares) and long-term (triangles) recovery scenarios of Table~\ref{tab:repair_times}. The BF is blue colored, while the BI is violet colored. Vertical separation within each configuration captures the dominant effect of recovery rate on $\rho$, while horizontal separation reflects the differential seismic reliability between the two archetypes. The four performance quadrants follow the same interpretation as Figure~\ref{fig:beta-rho-map}.}
    \label{fig:performance-space-spif}
\end{figure}

The joint $(\beta_{50y},\rho)$ performance space for the two industrial case studies is shown in Figure~\ref{fig:performance-space-spif}. Each structural configuration is represented by a set of markers corresponding to the short- and long-term recovery scenarios defined in Table~\ref{tab:repair_times}. The BF occupies the lower portion of the diagram ($\beta_{50}\approx 1$, $\rho\in[1.49,\,2.50]$), reflecting the combined effect of its relatively high seismic vulnerability and the slower recovery times associated with conventional steel construction. As noted in Section~\ref{sub:reliability-discussion}, the low $\beta_{50}$ of the BF is partly a consequence of its experimental design origin: the frame was conceived for shake-table testing campaigns aimed at studying the coupling between structural response and non-structural industrial components~\cite{bib:nardin2022BF, bib:nardin2022thesis}, not for life-cycle reliability performance. The BI system, by contrast, is located in the upper-right region of the diagram ($\beta_{50}\approx 2.08$, $\rho\in[3.15,\,4.16]$), confirming its superior performance along both dimensions simultaneously.
 
The nearly vertical separation between the short- and long-term recovery markers within each configuration illustrates that improvements in repair efficiency translate primarily into resilience gains with negligible impact on reliability, consistently with the trajectory structure discussed in Section~\ref{sec:toy-example}. The horizontal separation between the BF and BI configurations, on the other hand, reflects the difference in seismic fragility and is essentially unaffected by the recovery rate. Together, these observations reinforce the key design implication already identified in the analytical example: structural hardening (i.e., reducing the damage and collapse rates $\lambda_{01}$, $\lambda_{02}$, $\lambda_{12}$) is required to improve reliability, while investments in faster repair protocols (higher $\mu_{10}$) yield large gains in resilience with negligible impact on $\beta$.

\section{Conclusions and future directions}\label{sec:conclusions}
This study presents a continuous-time Markov chain (CTMC) formulation that generalizes the classical PEER-PBEE framework by embedding damage accumulation
and recovery directly into the stochastic evolution of the structural state, thereby relaxing the Poissonian assumption of perfect renewability after seismic events. The infinitesimal generator matrix $\bm Q$ unifies seismic hazard, state-dependent fragility, and recovery rates within a single operator, from which both reliability and resilience metrics are derived in a computationally efficient and physically interpretable manner.

A key finding of the industrial-scale applications is the fundamental  decoupling between reliability and resilience: long-term failure probabilities and reliability indices $\beta$ are governed almost exclusively by the seismic damage-to-collapse rates $\lambda_{02}$ and $\lambda_{12}$, and are largely insensitive to the recovery rate $\mu_{10}$ across the full range of plausible repair scenarios. Resilience, by contrast, is strongly controlled by the competition between damage accumulation and recovery, as reflected in the conditional occupation fraction $\mathcal{Q}_0$ and the standardized resilience index $\rho$. This decoupling, first identified analytically in the three-state example through the $(\beta,\rho)$ performance space, is confirmed quantitatively in both the braced frame and the base-isolated case studies. For the base-isolated system, the separation of recovery and damage timescales is so pronounced that $\rho$ remains near its maximum regardless of the recovery scenario, whereas for the braced frame a non-negligible sensitivity to $\mu_{10}$ persists, reflecting the less complete timescale separation in that configuration.

The spectral analysis of the transient generator $\bm Q_T$ provides an additional layer of insight: the dominant eigenvalue $\sigma_1$ sets the characteristic timescale of absorption, the quasi-stationary distribution $\bm q$ describes the conditional damage state occupancy prior to collapse, and the asymptotic resilience $\mathcal{R}_0^\infty = 1 - w_0 q_0$ links the macroscopic resilience metric directly to these spectral quantities. The large spectral ratios $|\sigma_2/\sigma_1|$ observed in both case studies ($10^2$--$10^5$) confirm that the quasi-stationary approximation is accurate well within standard engineering time horizons, making the spectral approach a practical and transparent tool for resilience assessment.

The proposed framework preserves the modular and non-intrusive nature of PBEE: existing fragility functions and hazard curves are embedded without modification, and recovery is introduced as a complementary rate-based component within the same architecture. Several extensions are natural within this setting. Time-varying generators driven by clustered seismicity models, such as the Epidemic-Type Aftershock Sequence (ETAS) model, would extend the formulation to non-homogeneous seismic input, particularly relevant for systems subject to rapid post-mainshock damage accumulation. The state-space representation also extends naturally to networked and portfolio-level assessments, enabling joint reliability and resilience quantification at the infrastructure scale. Future work will explore these directions, further consolidating the role of generalized PBEE formulations in supporting resilient design and management of civil infrastructure under realistic seismic hazard scenarios.

\section*{Acknowledgement \& Funding}
This research is supported by the Marie-Sklodowska Curie program and the REACTIS project, GA no. 101147351.
Views and opinions expressed are however those of the authors only and do not necessarily reflect those of the European Union, or REA or any sponsor. Neither the European Union nor the granting authority can be held responsible for them.
\newpage
\bibliography{bibliography.bib}

\appendix
\newpage
\renewcommand\thefigure{\thesection\arabic{figure}}
\setcounter{figure}{0} 

\section{- Appendix --  CTMC-based modelling: theoretical basis}\label{app:ctmc-basics}
Assuming the Markov property (Hyp \ref{hyp:memory}), the future evolution of the system depends only on its present state and not on its history. A stochastic process $DS(t)$ is said to be a CTMC if, for any pair $u_1 <u_2 , u_{1,2}\geq 0$, it satisfies:
\begin{equation} \label{eq:Markovprop}
\bP(DS(u_2) = ds_j \mid DS(u_1) = ds_i) = p_{ij}(u_1, u_2), \tag{A.1}
\end{equation}
where $p_{ij}(u_1, u_2)$ is the conditional probability of transitioning from state $ds_i$ at time $u_1$ to state $ds_j$ at time $u_2$. The collection of all such probabilities forms the transition probability matrix $\bm P(u_1, u_2) = \{p_{ij}(u_1, u_2)\}$.

If the process is homogeneous in time (Hyp~\ref{hyp:stat}), the transition probabilities depend only on the time lag $\Delta t=u_2-u_1$. In this case, we write:
\begin{equation} \label{eq:kernel}
    p_{ij}(u_1,u_2) = p_{ij}(u_2-u_1) := p_{ij}(\Delta t), \quad\quad \bm P(u_1,u_2):= \bm P(\Delta t). \tag{A.2}
\end{equation}
Under this condition, the evolution of the state probabilities is governed by the Chapman-Kolmogorov forward equations:
\begin{align}
    \bm \pi(t) &= \bm \pi(0) \bm{P}(t), \label{eq:CK1} \tag{A.3} \\
    \bm P(t+\delta t) &= \bm P(t) \bm P(\delta t), \label{eq:CK2} \tag{A.4}
\end{align}
where $\bm{\pi}(0)$ is the initial state distribution and $\bm P(0) =\bm I$ is the identity matrix.

To describe the instantaneous transition dynamics, we formally introduce the infinitesimal generator or jump rate matrix $\bm Q$ as:
\begin{equation}
    \bm{Q} = \lim_{\Delta t \to 0} \frac{\bm{P}(\Delta t) - \bm{I}}{\Delta t}. \tag{A.5}
\end{equation}

Each off-diagonal element $q_{ij}$ of $\bm Q$ represents the transition rate from state $ds_i$ to $ds_j$, while each diagonal entry is defined such that rows sum to zero: $q_{ii} = -\sum_{j \neq i} q_{ij}$. This matrix encapsulates the essential structure of the CTMC and governs the Kolmogorov forward equations, which describe how transition probabilities evolve over time:
\begin{align}
    \dot{\bm{P}}(t) &= \bm{P}(t) \cdot \bm{Q}, \quad \bm{P}(0) = \bm{I}. \label{eq:KDF1} \tag{A.6}
\end{align}

\begin{figure}[!ht]
    \centering
    \includegraphics[width=0.85\linewidth]{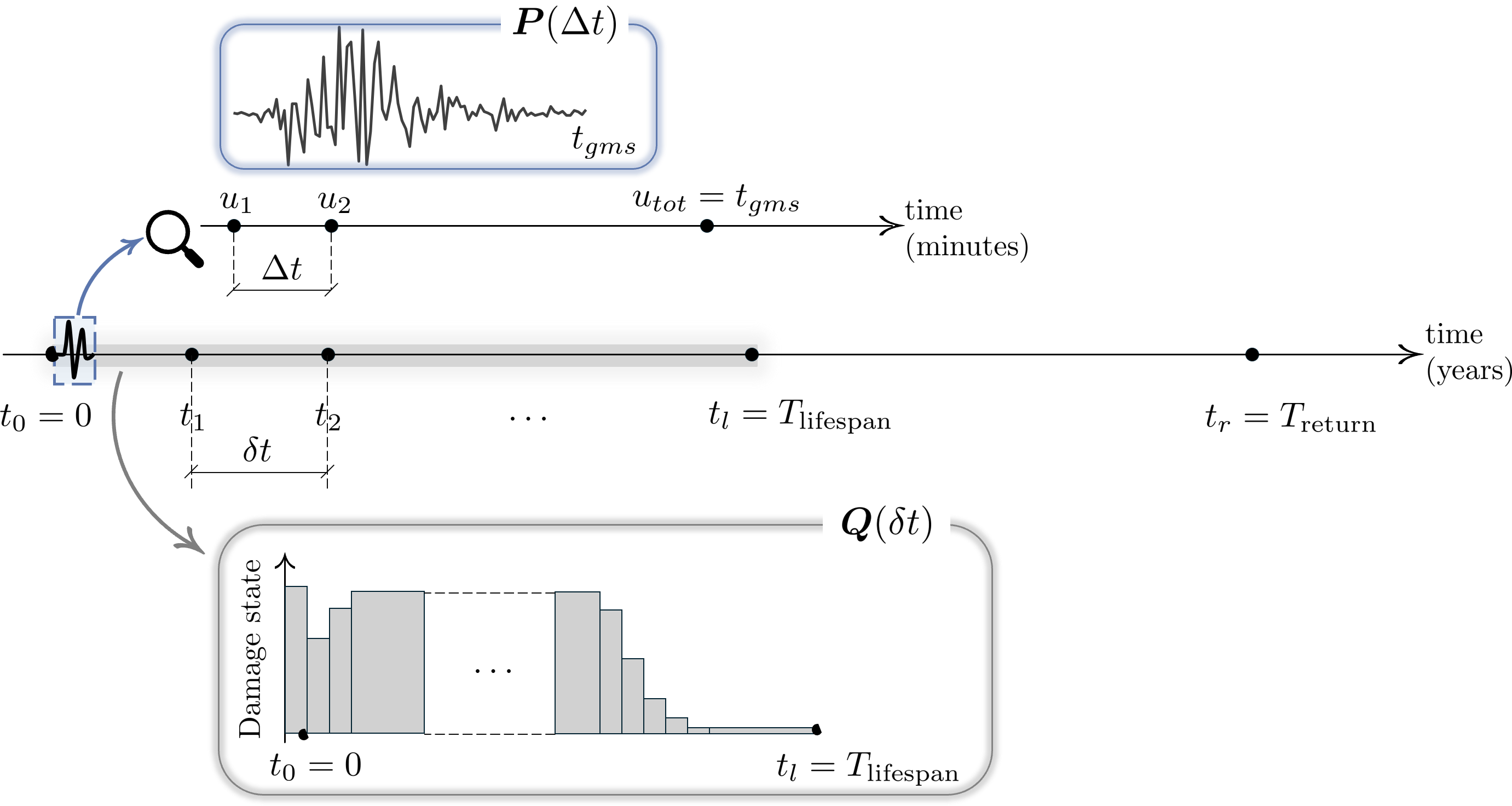}
    \caption{Overview and notation of the CTMC process. The upper axis shows the intra-event scale (seconds–minutes), where the transition probability kernel $\bm P(\Delta t)$ governs state transitions within a single ground motion. 
    The lower axis illustrates the inter-event scale (months–years), where the Kolmogorov equations, driven by the generator $\bm Q$, capture the long-term evolution of damage states over the system lifespan $T_{\text{lifespan}}$ and return period $T_{\text{return}}$ of the hazard.}
    \label{fig:timescales}
\end{figure}

Figure~\ref{fig:timescales} provides a schematic overview of the stochastic process $DS(t)$ and its time scale. 
The upper part illustrates the short-term, intra-event dynamics: within the duration of a single ground motion record of length $t_{gms}$, the system experiences rapid state transitions over time lags $\Delta t$ (order of seconds to minutes). 
These transitions are captured by $\bm P(\Delta t)$ as described in Eqn.~\eqref{eq:kernel}, which encodes the conditional probabilities of moving between damage states within an event. Strictly speaking, $\bm P(\Delta t)$ is the transition probability kernel of the CTMC, assigning conditional probabilities of moving between states over a time increment $\Delta t$.

In contrast, the lower part highlights the long-term, inter-event evolution of the process across the lifespan $T_{\text{lifespan}}$ of the structure and the return period $T_{\text{return}}$ of the hazard. 
Here, the relevant increment $\delta t$ is of the order of months or years, corresponding to the spacing of earthquake events or inspection/repair intervals. 
At this scale, the dynamic evolution and behavior of the system is governed by the Kolmogorov forward Eqns.~\eqref{eq:KDF1}--\eqref{eq:KDF2}, driven by the infinitesimal generator $\bm Q$. Indeed, differentiating \eqref{eq:CK1} and substituting \eqref{eq:KDF1}, we arrive at the system of Kolmogorov differential equations for state probabilities:
\begin{align}\label{eq:KDF2b}
    \dot{\bm{\pi}}(t) &= \bm{\pi}(t) \cdot \bm{Q}, \quad \bm{\pi}(0) = \bm{\pi}_0. \tag{A.7}
\end{align}
This equation defines a classical initial value problem whose solution yields the occupancy probabilities of each state at any time $t$. The general solution is:
\begin{align}\label{eq:KDF2_sol}
    \bm{\pi}(t) &= \bm{\pi}(0) \cdot \exp(\bm{Q} t). \tag{A.8}
\end{align}
Several analytical and numerical methods exist for solving Eqn.~\eqref{eq:KDF2b}, depending on the complexity of $\bm Q$; see \cite{bib:trivedi2017reliability, bib:Ross2014chapter4} for comprehensive overviews.

Here, $\bm P(t) = e^{\bm Q t}$ denotes the full transition probability matrix over all states. When the state space is partitioned into transient states $\mathcal{D}_T$ and an absorbing state $ds_C$ (see Eq.~\eqref{eq:Q_QT}), the transient-to-transient block of $\bm P(t)$ reduces to $\bm P_{TT}(t) = e^{\bm Q_T t}$, which governs transitions among operational states prior to absorption.
\FloatBarrier

\section{- Appendix --  Derivation of $\mathbf{P_{TT}(t)}$}\label{app:ctmc-ptt}
\setcounter{figure}{0} 
The generator $\bm Q$ partitions as in Eq.~\eqref{eq:Q_QT}. Since $\bm Q$ is block upper-triangular, the transition matrix $
\bm P(t)=e^{\bm Qt}$ inherits the same structure:
\begin{equation}\label{eq:C01-Pt-block}
\bm P(t)=
\begin{bmatrix}
\bm P_{TT}(t) & \bm p_{TA}(t)\\
\bm 0^\intercal & 1
\end{bmatrix}, \tag{B.1}
\end{equation}
where $p_{AA}(t)=1$ follows immediately from the fact that the absorbing state cannot be exited, and
\begin{equation}
[\bm P_{TT}(t)]_{ij}=p_{ij}(t),
\qquad i,j\in\mathcal{D}_T. \notag    
\end{equation}

Substituting \eqref{eq:C01-Pt-block} into the Kolmogorov forward equation \eqref{eq:KDF1} and matching blocks gives
\begin{equation}\label{eq:C02-PTT-ode}
\dot{\bm P}_{TT}(t)=\bm Q_T\bm P_{TT}(t),
\qquad
\bm P_{TT}(0)=\bm I_C, \notag
\end{equation}
with solution
\begin{equation}
\bm P_{TT}(t)=e^{\bm Q_T t}, \notag    
\end{equation}
and
\begin{equation}\label{eq:C03-pTA-ode}
\dot{\bm p}_{TA}(t)=\bm Q_T\bm p_{TA}(t)+\bm a,
\qquad
\bm p_{TA}(0)=\bm 0. \tag{B.2}    
\end{equation}
Applying variation of constants to \eqref{eq:C03-pTA-ode} yields
\begin{equation}\label{eq:C04-pTA-sol}
\bm p_{TA}(t)=\int_0^t e^{\bm Q_T(t-s)}\bm a\,ds. \tag{B.3}
\end{equation}
Alternatively, since every row of $\bm P(t)$ sums to one, \eqref{eq:C04-pTA-sol} simplifies to
\begin{equation}\label{eq:C05-pTA-simple}
\bm p_{TA}(t)=\bm 1_T - e^{\bm Q_T t}\bm 1_T. \notag
\end{equation}
The full transition matrix is therefore
\begin{equation}\label{eq:C06-Pt-full}
\bm P(t)=
\begin{bmatrix}
e^{\bm Q_T t} & \bm 1_T - e^{\bm Q_T t}\bm 1_T\\[4pt]
\bm 0^\intercal & 1 \notag{B.5}
\end{bmatrix}.
\end{equation}

\section{- Appendix --  Derivation of the expected occupation time $ds_0$}\label{app:ctmc-a0}

Starting from the definition in Eq.~\eqref{eq:occupation_time_0} and 
applying Fubini's theorem to exchange expectation and integration,
\begin{equation}\tag{C.1}\label{eq:C1}
\mathbb{E}[A_0(t)\mathbf{1}_{\{\mathcal T>t\}}]
=
\int_0^t \mathbb{P}(DS(s)=ds_0,\;\mathcal T>t)\,ds.
\end{equation}
The joint probability decomposes via the Markov property as
\begin{equation}\notag
\mathbb{P}(DS(s)=ds_0,\;\mathcal T>t)
=
\underbrace{\bm e_0^\intercal e^{\bm Q_T s}\bm e_0}_{\mathbb{P}(DS(s)=ds_0)}
\cdot
\underbrace{\bm e_0^\intercal e^{\bm Q_T(t-s)}\bm 1_T}_{\mathbb{P}(\mathcal T>t\mid DS(s)=ds_0)}.
\end{equation}
Introducing the selector $\bm D_0 := \bm e_0\bm e_0^\intercal$, the two factors 
combine into a single matrix product, and substitution into 
Eq.~\eqref{eq:C1} gives
\begin{equation}\tag{C.2}\label{eq:C2}
\mathbb{E}[A_0(t)\mathbf{1}_{\{\mathcal T>t\}}]
=
\int_0^t
\bm e_0^\intercal\,e^{\bm Q_T s}\,\bm D_0\,e^{\bm Q_T(t-s)}\,\bm 1_T\,ds,
\end{equation}
which is Eq.~\eqref{eq:conditional_mean_occupation_numerator} of the main text.

\section{- Appendix -- Asymptotic limit of $\mathcal{R}_0(t)$}
\label{app:ctmc-R0-asymptotic}

Starting from Eq.~\eqref{eq:Q0_def} and using Eq.~\eqref{eq:C2}, the complement of $\mathcal{R}_0$ reads
\begin{equation}\tag{D.1}\label{eq:D1}
1-\mathcal{R}_0(t)
=
\frac{1}{t\,S_{\mathcal T}(t)}
\int_0^t
S_{\mathcal T}(s)\;\widetilde{\pi}_0(s)\;g_0(t-s)\,ds,
\end{equation}
where $\widetilde{\pi}_0(s) = \pi_0(s)/S_{\mathcal T}(s)$ is the conditional probability of occupying $ds_0$ given survival, and $g_0(u) = \bm e_0^\intercal e^{\bm Q_T u}\bm 1_T$.

Assume $\bm Q_T$ is irreducible with dominant eigenvalue $\sigma_1 < 0$, right eigenvector $\bm w$, and left eigenvector $\bm \nu$ (the QSD), normalized as in Eq.~\eqref{eq:qsd_eigenvectors_main}. 
The rank-one spectral asymptotics of $e^{\bm Q_T t}$ give
\begin{equation}\tag{D.2}\label{eq:D2}
S_{\mathcal T}(t) \sim w_0\,e^{\sigma_1 t},
\qquad
g_0(t) \sim w_0\,e^{\sigma_1 t},
\qquad t\to\infty,
\end{equation}
with $w_0 = \bm e_0^\intercal\bm w$, and $\widetilde{\pi}_0(t)\to\nu_0$ 
as established in Eq.~\eqref{eq:tilde_pi0_limit_main}. Substituting 
Eq.~\eqref{eq:D2} into Eq.~\eqref{eq:D1}, the exponential factors cancel 
between numerator and denominator, leaving
\begin{equation}\tag{D.3}\label{eq:D3}
1-\mathcal{R}_0(t)
=
\frac{w_0}{t}\int_0^t \widetilde{\pi}_0(s)\,ds
+O\!\left(\tfrac{1}{t}\right),
\qquad t\to\infty.
\end{equation}
Since $\widetilde{\pi}_0(s)\to\nu_0$, Cesàro convergence applied to 
Eq.~\eqref{eq:D3} yields
\begin{equation}\tag{D.4}\label{eq:D4}
\lim_{t\to\infty}\mathcal{R}_0(t) = 1 - w_0\,\nu_0,
\end{equation}
which is Eq.~\eqref{eq:R0_limit_complement_main} of the main text.

\setcounter{figure}{0} 
\section{- Appendix --  Supplementary material for the industrial case study}\label{app:spif}

The seismic hazard input adopted in this work refers to the municipality of L'Aquila, central Italy, one of the most seismically active regions in Europe. Hazard data are extracted from the European Seismic Hazard Model (ESHM20) \cite{bib:danciu2021eshm20}, available through the EFEHR portal,\footnote{\url{https://hazard.efehr.org/en/hazard-data-access/hazard-curves/}} and correspond to a 10\% probability of exceedance (PoE) in 50 years, equivalent to a return period of $T_{\text{return}} \approx 475$ years.

\begin{figure}[!ht]
    \centering
    \includegraphics[width=0.75\linewidth]{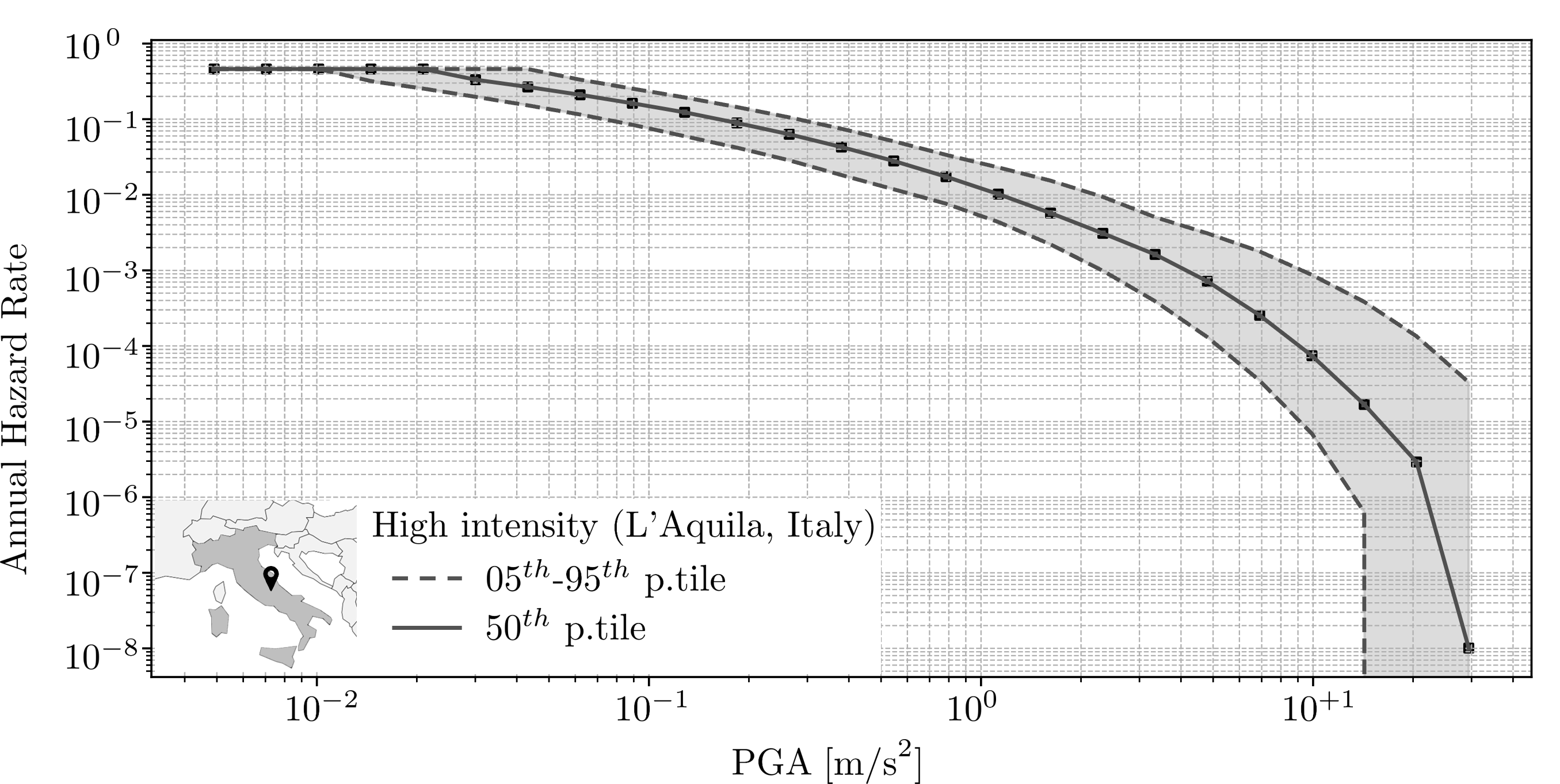}
    \caption{Seismic hazard curve of L’Aquila, central Italy, retrieved from the ESHM20 model \cite{bib:danciu2021eshm20}. Dotted lines represent the $5^{th}$ and $95^{th}$ percentiles, while the solid line shows the $50^{th}$, corresponding to a $10 \%$ PoE in 50 years.}
    \label{fig:seismic-hazard}
\end{figure}

Figure~\ref{fig:seismic-hazard} shows the geographic location of the site and the associated seismic hazard curve, including the median annual exceedance rates alongside the 5$^{\text{th}}$ and 95$^{\text{th}}$ percentile bounds, which reflect epistemic uncertainty in the underlying ground motion models.

\end{document}